\documentstyle[12pt,epsf,rotate]{article}
\def\ba{\begin{array}}
\def\ea{\end{array}}
\def\bc{\begin{center}}
\def\ec{\end{center}}
\def\be{\begin{equation}}
\def\ee{\end{equation}}
\def\bfe{\begin{figure}[h]}
\def\efe{\end{figure}}
\def\btr{\begin{tabular}}
\def\etr{\end{tabular}}
\def\bea{\begin{eqnarray}}
\def\eea{\end{eqnarray}}

\def\sun{\hbox{$\odot$}}
\def\farcs{\hbox{$.\!\!^{\prime\prime}$}}

\newcommand{\abtw}{2 \sqrt{1-2\beta}\;\afa\frac{y_{ls}}{y_{os}}}
\newcommand{\acr}{a_{\rm cr}}
\newcommand{\afa}{\alpha_{0}}
\newcommand{\am}{A_{\rm M}}
\newcommand{\Bave}{<\!\!B(<m\!)\!\!>}
\newcommand{\bt}{\rm B_{T}}
\newcommand{\dmin}{\Delta_{\rm min}}

\newcommand{\dtheta}{\Delta \theta}

\newcommand{\hnot}{{\rm H_{o}}}
\newcommand{\lco}{L_{\rm o}}
\newcommand{\llstar}{\frac{L}{L^{*}}}
\newcommand{\mbstar}{\rm M_{B_{T}\left( 0\right) }^{*}}
\newcommand{\mmin}{m_{\rm min}}
\newcommand{\mo}{m_{\rm o}}
\newcommand{\nexp}{N_{\exp}}

\newcommand{\nqdm}{N_{\rm Q}(m+\Delta)}
\newcommand{\nqm}{N_{\rm Q}(m)}
\newcommand{\om}{\Omega_{\rm o}}
\newcommand{\omda}{\Omega_{\Lambda}}
\newcommand{\omr}{\Omega_{\rm R}}
\newcommand{\pde}{p(\Delta)}
\newcommand{\rc}{r_{\rm c}}

\newcommand{\sigstardm}{\frac{\vdm}{\vdm^{*}}}
\newcommand{\thetaq}{\theta_{\rm Q}}
\newcommand{\yy}{\frac{y_{ls}}{y_{os}}}
\newcommand{\bj}{\rm B_{J}}

\newcommand{\lll}{\log_{10}\left( L/L^{*}\right)}

\newcommand{\lv}{\log_{10}(\sigma}

\newcommand{\re}{R_{\rm e}}

\newcommand{\vdm}{\sigma_{\rm DM}}

\newcommand{\vlos}{\sigma_{\rm los}\left( R\right)}
\newcommand{\vsre}{\sigma_{\re}^{*}}
\newcommand{\vvratio}{\sigma_{\rm los}^{2}/\vdm^{2}}

\begin{document}

\rightline{\bf CWRU-P30-98}

\bc
{\Large\bf
GRAVITATIONAL LENSING STATISTICS AND CONSTRAINTS ON THE
COSMOLOGICAL CONSTANT REVISITED}

\vskip 14pt
Yu-Chung N. Cheng\footnote{E-mail address: yxc16@po.cwru.edu}
and
Lawrence M. Krauss\footnote{Also Dept. of Astronomy, E-mail address:
krauss@theory1.phys.cwru.edu}

\vskip 14pt
\small\it 
Department of Physics, Case Western Reserve University, Cleveland, 
OH 44106-7079

\ec

\begin{quote}
We re-analyze constraints on the cosmological constant that can be obtained
by examining 
the statistics of strong gravitational lensing of distant quasars by
intervening galaxies, focusing on 
uncertainties in galaxy models (including
velocity dispersion, luminosity functions, core radii and magnification bias
effects)  and on the parameters of the galaxy distribution and luminosity
functions. In the process we derive new results on magnification biasing for
galaxy lenses with non-zero core radii, and on how to infer the proper
velocity dispersions appropriate for use in 
lensing statistics.  We argue that the
existing data do not disfavor a large cosmological constant.  In fact, for a
set of reasonable parameter choices, using the results of 5 optical quasar
lensing surveys we find that a maximum likelihood analysis favors a value of
$\om$ in the range $ \approx 0.25$-0.55 
in a flat universe. An open cosmology is not favored
by the same statistical analysis.  Systematic uncertainties
are likely to be dominant, however, as these
results are sensitive to uncertainties in our understanding of galaxy
luminosity functions, and dark matter velocity dispersions, as well as
the choice of lensing survey, and to a lesser extent the existence of
core radii. Further observational work will be required before it is
possible to definitively distinguish between cosmological models on the
basis of gravitational lensing statistics.
\end{quote}


\section{Introduction}

Recently there has been a renewed interest in the possibility that the
cosmological constant may dominate the energy density of the universe, in order
to resolve several cosmological conundra
\cite{KT,oststein,krauss}, and more recently due to direct observational
measurements of the distance redshift relation \cite{perl,riess}.  For some
time, one of the few cosmological pieces of evidence which has appeared
apparently disfavoring this possibility has involved analyses of the statistics
of gravitational lensing of quasars by galaxies
\cite{FT,KW,Kochanek93,Kochanek96}.

The effort to constrain cosmological parameters utilizing such statistics is of
course highly dependent both on the quality of the existing data, and also on
the robustness of the theoretical inputs in the analysis.  Because of the
potential importance of the claimed constraints on a cosmological constant, it
is worth re-examining in some depth the dependence of the resulting constraints
on model dependent assumptions, as well as making some attempt to improve the
models of galaxies one uses in the analysis.  

In this paper, we re-analyze lensing statistics, concentrating on the
role of the luminous E/S0 galaxy distribution function and 
luminosity function parameters, and the self consistent
modelling of galaxy parameters, including core radii. As we shall show,
new considerations of the velocity dispersions of lensing  galaxies
\cite{Kochanek93,Kochanek94}, the core radii of lensing galaxies \cite{KW,HK},
and the magnification bias resulting from selection effects 
\cite{Kochanek93,Kochanek96,HK,TOG} both affect the nature of the
constraints one derives, and also imply that the constraints presently possible
from the statistics of strong lensing may vary over a wide range. Nevertheless,
we find for reasonable parameters, a best fit cosmology with a low matter 
density flat universe with a cosmological constant~\cite{cy}\footnote{As this
paper was being finalized for submission a new investigation~\protect\cite{cy2}
also appeared which explores several similar issues and reaches compatible
conclusions.}.  Moreover, we find that the data that favors a low
$\om$ mildly favors a flat cosmological constant dominated universe over an open
universe.  Five sets of quasar surveys are considered in order to compare
predictions and observations.  

\section{Theory layout: core radii, magnification bias, lensing statistics,
and galaxy parameters}

In this section we first review the basic formalism we shall utilize in our
statistical analyses, including the calculation of optical
depths for lenses modelled as isothermal spheres with core radii. We then
proceed to a new calculation  of the relevant magnification
bias which should be utilized in this situation.  This is significant, because
the trade off between reduction in lensing probability caused by finite
cores is offset to some degree 
by an increase in the magnification bias due to lensing \cite{Kochanek96}.  The
degree of this offset is important, however, if the two effects do not
precisely cancel, as we find.  We conclude this section with a short summary
of our statistical maximum likelihood method.

\subsection{Optical depths and nonsingular isothermal spheres}

We begin by modelling the 
mass density distribution of elliptical galaxies with the
following form \cite{HK}:
\be
\rho (r) = \frac{\sigma_{\rm DM}^2}{2 \pi G (r^{2}+\rc^{2})}
\label{eq:density}
\ee
where $\sigma_{\rm DM}$ is the velocity dispersion of this system (presumably
the dominant dark matter, a consideration we shall return to later), and
$\rc$ is the core radius. When $\rc$ is zero, this model reverts to a singular
isothermal sphere (SIS) model. With the density distribution from
equation~(\ref{eq:density}), one
can calculate the lensing cross section, $\sigma_{\rm cs}$, within which
multiple images of gravitational lensing events are observed.
\be
\sigma_{\rm cs} = \pi \acr^{2} f(\beta)
\ee
where
\be
\acr =
\frac{c\,\afa\,y_{o\ell}\,y_{\ell s}}{\hnot (1+z_{\ell}) y_{os}}
\ee
$\acr$ is the critical radius, $z$ is the redshift, and 
$\afa = 4 \pi (\frac{\vdm}{c})^{2}$, is the bend angle for
an isothermal sphere with core radius $\rc = 0$. The Hubble constant $\hnot$ is
100 $h$ km s$^{-1}$ Mpc$^{-1}$.
(Please note that this $\acr$ was the size
at lens redshift $z_{l}$. It is not the critical radius at the present
Universe.) The $y$ quantities are angular
diameter distances between the source, lens and observer, and will be discussed
in  more detail below.  The quantity,
$\beta \equiv \rc/\acr$ as defined in Hinshaw \& Krauss \cite{HK}, and
\be
f(\beta) \equiv 1 + 5\beta - \frac{1}{2} \beta^2 - \frac{1}{2}
\sqrt{\beta}(\beta + 4)^\frac{3}{2}
\label{eq:fb}
\ee
These quantities arise in the determination of 
the bend angle, $\alpha$, of the light trace from the source,
which in the case of an isothermal sphere with a finite core radius,
is a function of the velocity dispersion $\vdm$, impact
parameter $b$, and core radius $\rc$ \cite{HK}. The general formula for the 
bend angle is:

\be
\alpha (b) = \frac{4 b}{c^{2}} \int_{b}^{\infty} dr \frac{\partial \Phi}
{\partial r} \frac{1}{\sqrt{r^{2} - b^{2}}}
\ee
where
\be
\Phi (\vec{r}) = - \int_{v} d^{3}\vec{r^\prime} \frac{G \rho (\vec{r^\prime})}
{\left| \vec{r} - \vec{r^\prime} \right|}
\ee
After calculating bend angles of multi images, one can then calculate the image
separation, $\Delta\theta$, and compare to measurements.
 
The full lensing
probability can then also be calculated as follows:
\be
\tau  =  \int d\tau  
 =  
\int_{0}^{z_{s}} d z_{l}\;\int_{0}^{\infty}d\left(\llstar\right) 
\left(\frac{y_{ol}y_{ls}}{y_{os}} \right) ^2
\frac{F B(<m) f(\beta)}{\sqrt{\om (1+z_{l})^{3} + \omr (1+z_{l})^{2} + \omda}}
\left(\llstar\right)^{\alpha} \exp \left(-\llstar\right) 
\label{eq:prob}
\ee
where
\be
F = \frac{c^{3} \pi n \afa^{2}}{\hnot^{3}}.
\label{eq:ffactor}
\ee
The galaxy number 
density $n$ is the luminous E/S0 galaxy density, which is about 30 per cent of
the total galaxy number density \cite{pg}, which is taken to be
is $1.40\times 10^{-2} h^3$ Mpc$^{-3}$, given by Loveday et al. 
\cite{Loveday}.
Spiral galaxies are not important in studies of strong
lensing, because of their huge core radii \cite{KW}.
$\omr$ is the curvature term, and $\omda$ is the cosmological constant, so
that $\om + \omr + \omda = 1$ in all cases.
In a flat universe, $y_{ls}$ is simply $y_{os}$-$y_{ol}$. However, in
an open universe (with $\omda =0$):
\be
y_{ls} = y_{os}\sqrt{1+\omr y_{ol}^{2}}-y_{ol}\sqrt{1+\omr y_{os}^{2}}
\ee
$y_{oi}$ 
is the angular size distance \cite{Peebles}:
\be
y_{oi} \equiv 
\frac{1}{\sqrt{\omr}}\sinh\left(\sqrt{\omr}
\int_{0}^{z_i} \frac{dz}{\sqrt{\om (1+z)^{3} + \omr (1+z)^{2} + \omda}}\right)
\label{eq:ydist}
\ee
$B(<m)$ is the magnification bias factor, which takes into account that
lensed quasars are magnified, and thus have a larger probability of being
observed than unlensed quasars, as we shall discuss in the next section. 
We also have to integrate the lensing 
probability
density over the lens luminosity distribution $L$, which is assumed to be given
by
the Schechter luminosity function \cite{schechter}:
\be
\phi(L) dL = \left(\frac{L}{L^{*}}\right)^{\alpha}
\exp \left(-\frac{L}{L^{*}}\right)
d\left(\frac{L}{L^{*}}\right)
\label{eq:schechter}
\ee
The parameters of the Schechter function will be discussed in some detail
later, as they induce perhaps the dominant uncertainties in the
predictions for the gravitational lensing of quasars by galaxies.

\subsection{Magnification bias}

Magnification bias in a lensing analysis is due to the amplification 
of lensed quasar images. Because of this effect, an observer can observe
a lensing event which cannot be seen if either it is not lensed or its 
magnification factor is not large enough to raise it above the minimum
sensitivity of a flux-limited survey. A magnification bias factor, which
enhances the optical depth compared to the bare optical depth calculated in
the absence of this selection effect, can be estimated based on flux limits
and the presumed luminosity distribution of quasars.  This factor has been
estimated earlier to be as large as 26 in the case of initial lensing surveys
using a singular isothermal sphere (SIS) model for galaxy lenses
\cite{TOG}. Its formula and its value have been re-examined by Fukugita
\& Turner 
\cite{FT}, and the value
can be as low as 4, if the survey of a complete set of quasars with apparent 
magnitude less than 22 can be achieved. 
Amplification of a lensed quasar including a core
radius of the lensing galaxy has been discussed in Hinshaw \& Krauss 
\cite{HK}. However, the bias
factor has not been calculated before for this case.  Here we carry out such an
analysis.

We begin with the definition of amplification ($A$) of a lensed quasar and also 
include the core radius in our formulas. Following Peebles \cite{Peebles},
\be
A = \left| \frac{\theta d\theta}{\thetaq d\thetaq} \right|
= \left| \frac{b db}{\ell d\ell} \right|
= \left| \frac{x dx}{L_1 dL_1} \right|
\label{eq:first}
\ee
where $\theta$ is the angle between the lens and the pseudo image of the quasar,
$\thetaq$ is the angle between the lens and the real quasar, $b$ is the impact
parameter, $\ell$ is the transverse (projected) distance of the lens center from
the line of sight, and $x \equiv b/\acr, L_1 \equiv \ell/\acr$.  
We will see that the amplification $A$ is a function of
$L_1$ and $\beta$ only, as discussed below. Following equation~(6) in 
Hinshaw \& Krauss \cite{HK},
$0 \leq L_1 \leq \lco \equiv \frac{\ell_{0}}{\acr} \leq 1$, and
$0 \leq \beta \leq \frac{1}{2}$, we can easily rewrite that equation to be
\be
L_1 = -x + \frac{\sqrt{x^{2} + \beta^{2}} - \beta}{x}
\label{eq:lxb}
\ee
This means that although $A$ is a function of $L_1, x,$ and $\beta$, we can
rewrite $A$ in terms of $L_1$ and $\beta$ only. Using
equation~(\ref{eq:lxb}), we have
\be
\frac{dL_1}{dx} = \frac{1}{\sqrt{x^{2} + \beta^{2}}} - \frac{L_1}{x} - 2
\ee
and we can calculate the amplification $A$. For the total amplification, we sum 
up the absolute values of amplification caused by each individual image, i.e.,
\be
A = \sum_{i} \left| \frac{x_{i} dx_{i}}{L_1 dL_1} \right|
\ee
where $x_{i}$ are the solutions of equation~(\ref{eq:lxb}), shown in 
Cheng \& Krauss \cite{darkL}. With
\be
\ell_{0}^{2} \equiv \acr^{2} + 5\acr \rc -\frac{1}{2} \rc^{2} - \frac{1}{2} 
\rc^{1/2} (\rc + 4 \acr)^{3/2}
\ee
(we should remind our readers that $\lco^2$ is actually identical to 
equation~(\ref{eq:fb}) above.) 
we can define the averaged amplification as a function of $\beta$:
\be
<\!A\!> \equiv \frac{\int^{\lco}_{0} dL_1\, L_1\, 
A(L_1,\beta)}{\int^{\lco}_{0} dL_1\, 
L_1} = \frac{2}{\lco^{2}} \int^{\lco}_{0} dL_1\, L_1\, A(L_1,\beta)
\label{eq:ampave}
\ee
The numerical values of $<\!A\!>$ are listed in Table~\ref{table:ampave}. 

Our next step is to calculate the minimum value of amplification ($\am$) from 
the three
quasar images for a given $\beta$, which will be used in the magnification bias
calculation. The numerical results are listed in Table~\ref{table:amin}. The 
values of $\beta$, the corresponding minimum values of amplification,
the corresponding $L_1$ values at $\am$, and the corresponding $\lco$ are listed
in Table~\ref{table:amin}. We have also found that $\am$ can be well fitted by
$2/(\lco^{0.65})$ and plotted in Figure~\ref{fig:amin}. We will use this 
approximation in our magnification bias calculation.

We can now derive the magnification bias. We start with the 
probability density of amplification
\be
p(A) dA = 2 \am^{2} A^{-3} dA, {\hskip 1in} A \geq \am
\ee
such that
\be
\int_{\am}^{\infty} p(A) dA = 1
\ee
If we define 
\bea 
\Delta & \equiv & 2.5 \log_{10} A \nonumber\\
\dmin  & \equiv & 2.5 \log_{10} \am
\eea
then we have
\be
p(A) dA = \pde d\Delta = 0.8 (\ln10) \am^{2} 10^{-0.8\Delta} d\Delta
\ee
The magnification bias is defined as:
\be
B(m) \equiv \frac{1}{\nqm} \int_{\dmin}^{\infty} \nqdm \pde d\Delta
\ee
where
\be
\nqm \propto \left\{ 
\ba{ll}
10^{a(m-\mo)}, & \mbox{if $m \leq \mo$}\\
10^{b(m-\mo)}, & \mbox{if $m \geq \mo$}
\ea
\right.
\label{eq:distri}
\ee
(Note the quantity $b$ above should not be confused with the quantity
used earlier to describe the impact parameter for lensing.)
We then calculate the magnification bias:
\be
B(m) = \left\{
\ba{ll}
0.8 \am^{2} \left[\left(\frac{1}{a-0.8}+\frac{1}{0.8-b}\right) 
10^{(a-0.8)(\mo-m)}
- \frac{1}{a-0.8}\am^{(2.5a-2)}\right] & \mbox{if $m \leq \mo-\dmin$} \\
\frac{0.8}{0.8-b}\am^{2.5b} 10^{(b-a)(m-\mo)} & 
\mbox{if $\mo-\dmin \leq m \leq \mo$} \\
\frac{0.8}{0.8-b} \am^{2.5b} & \mbox{if $\mo \leq m$}
\ea
\right.
\ee
Interestingly, $b$ has to be less than 0.8 from the above formula in order to
have a reasonable $B(m)$. The observational value of $b$ is well below 0.8 and
will be presented later.

Next, as discussed in Fukugita \& Turner \cite{FT}, we have to average 
$B(m)$ over the observed magnitude ($m$) distribution in order to obtain the 
relevant collective bias
$B(<m)$. We approximate a magnitude limited quasar survey using the
selection function:
\be
S(m) = \left\{ 
\ba{ll}
1, & \mbox{when $m$ is less (brighter) than the survey limit}\\
0, & \mbox{when $m$ is larger (dimmer) than the survey limit}
\ea
\right.
\ee
and define
\be
B(<m)  \equiv  \frac{\int_{\mmin}^{m} dm \int_{\dmin}^{\infty} d\Delta \pde \nqdm}
{\int_{\mmin}^{m} dm \nqm}
\label{eq:avebias}
\ee
Using equation~(\ref{eq:distri}), we find
\bea
B(<m) & = & \frac{a(a-b)}{(a-0.8)(0.8-b)} \left(
\frac{10^{0.8(m-\mo)}-10^{0.8(\mmin-\mo)}}
{10^{a(m-\mo)}-10^{a(\mmin-\mo)}}\right) \am^{2} - \frac{0.8}{a-0.8} \am^{2.5a} 
\nonumber \\
& & \mbox{when $\mmin \leq m \leq \mo-\dmin < \mo$} \nonumber \\
& & \nonumber \\
B(<m) & = & \frac{1}{10^{a(m-\mo)}-10^{a(\mmin-\mo)}} \left(
\frac{b-a}{b}+\frac{0.8}{a-0.8}10^{a(\mmin-\mo)}\am^{2.5a} \right. \nonumber \\
 & & \left. +\frac{0.8 a}{b(0.8-b)}10^{b(m-\mo)}\am^{2.5b} 
-\frac{a(a-b)}{(a-0.8)(0.8-b)} 10^{0.8(\mmin-\mo)}\am^{2} \right) \nonumber \\
& & \mbox{when $\mmin \leq \mo-\dmin \leq m \leq \mo$}\nonumber \\
& & \nonumber \\
B(<m) & = & \frac{0.8 a}{(0.8-b)b} \left(\frac{10^{b(m-\mo)}-10^{b(\mmin-\mo)}}
{10^{a(m-\mo)}-10^{a(\mmin-\mo)}}\right) \am^{2.5 b}\nonumber \\
& & \mbox{when $\mo-\dmin \leq \mmin \leq m \leq \mo$}\nonumber \\
& & \nonumber \\
B(<m) & = & \left(\frac{0.8}{0.8-b}\right)\frac{10^{b(m-\mo)}-10^{b(\mmin-\mo)}}
{\frac{b}{a}\left(1-10^{a(\mmin-\mo)}\right)+10^{b(m-\mo)}-1} \am^{2.5 b}
\nonumber \\
& & \mbox{when $\mo-\dmin \leq \mmin \leq \mo \leq m$}\nonumber \\
& & \nonumber \\
B(<m) & = & \frac{1}{b(1-10^{a(\mmin-\mo)})+a(10^{b(m-\mo)}-1)} \left(
b-a +\frac{0.8 b}{a-0.8} 10^{a(\mmin-\mo)}\am^{2.5a} \right. \nonumber \\
& & \left.  + \frac{0.8 a}
{0.8-b}10^{b(m-\mo)}\am^{2.5b}-\frac{ab(a-b)}{(a-0.8)(0.8-b)}10^{0.8(\mmin-\mo)}
\am^{2} \right) \nonumber \\
& & \mbox{when $\mmin \leq \mo-\dmin < \mo \leq m$} 
\eea

\subsection{Lensing statistics}

We utilize a maximum likelihood method based on Poisson statistics in our
analysis
\cite{loredo}.
If we have a complete and flux limited quasar
survey, then we can calculate the expected number of lensing events, $\nexp$, by
summing up the probability of each quasar over all quasar samples, $N_{\rm Q}$,
i.e.,
\be
\nexp = \sum_{i=1}^{N_{\rm Q}} \tau(z_{s}=z_{i})
\label{eq:explens}
\ee
Using equations~(\ref{eq:prob}) and~(\ref{eq:explens}), we can write down the 
expected number density 
distribution, ${\cal N}_i$, which is the integrand of equation~(\ref{eq:prob}).
If we choose small bin sizes for source
redshift, galaxy redshift, and luminosity $\llstar$, then there will be either 
no lensing events or one lensing event in each bin. Assuming appropriate 
Poisson 
probabilities $P_{0,i}$ and $ P_{1,i}$, respectively,
the  likelihood function is
\be
{\cal L} = \left[\prod_{i=1}^{N_{\rm obs}} P_{1,i}\right] 
\left[\prod_{i=1}^{N_{\rm un}} P_{0,i}\right]
\ee
where $N_{\rm obs}$ is the observed number of multiple-imaged lensing events, 
and $N_{\rm un}$ is
the number of the un-lensed events. After taking the logarithm of $\cal L$,
and taking the limit of summation to be an integral, we find the formula used 
in our lensing statistical analysis:
\be
\ln {\cal L} = \sum_{l=1}^{N_{\rm obs}} \ln {\cal N}_{l} - \nexp +{\rm constant}
\ee
with
\begin{eqnarray}
\ln{\cal N}_{l} & \equiv & \ln\frac{d^{2} \tau}{d z_{l}\;d\left(\llstar\right)}
 \\
& = & \ln\left[\left(\frac{\sigma_l}{\vdm^*}\right)^{4} 
\left(\frac{y_{ol}y_{ls}}{y_{os}} 
\right)^2 \frac{B(<m) f(\beta)}
{\sqrt{\om (1+z_{l})^{3} + \omr (1+z_{l})^{2} + \omda}}
\left(\frac{L_l}{L^*}\right)^{\alpha}\right] 
 - \left(\frac{L_l}{L^*}\right) + {\rm constant} \nonumber
\label{eq:Nevent}
\end{eqnarray}

\section{Velocity dispersions and luminosity functions: dark matter and
galaxies}

The probability of gravitational lensing can be seen, from equation
(\ref{eq:prob}) to depend strongly on both the luminosity function, and the
velocity dispersion of galaxies.  There is an explicit dependence on the latter
in the lensing cross section.  An implicit dependence arises as a result of the
need to relate velocity dispersion to luminosity when performing the integral
in equation (\ref{eq:prob}).  We have recently \cite{cheng} explored
the model dependent 
considerations
associated with extracting the relevant velocity dispersions of galaxies from
the data, in the context of finite core isothermal models.  We review
the chief results here, along with discussing 
the interdependence of these estimates on
the assumed luminosity distribution of galaxies.

Returning to equation (\ref{eq:density}), we should bear in mind that the 
velocity
dispersion in this equation, which is assumed to be independent of radius,
cannot be measured directly.
What we can measure is the line-of-sight velocity dispersion
$\vlos$, with the projected distance $R$ measured from the center of the 
observed galaxy (i.e., $R$ is perpendicular to the line-of-sight to us). 
For a singular isothermal sphere, $\vdm $ is not a function of $R$, but
$\vlos$ is, and this relationship is different when a galaxy has 
a finite core.  Thus, one
has to be careful how to use the measured velocity dispersion to derive the
relevant quantity to utilize for lensing, namely whether it well approximates
$\vdm$. Using a wide 
range of values of galaxy models, and self consistently solving the dynamical
equations we demonstrated \cite{cheng} 
that $\vvratio$ is strongly sensitive to
the core radius and galaxy anisotropy 
when the projected radius is less than 0.1$\re$. This is
consistent with observational data \cite{BSG}.
It can thus be dangerous to simply
consider the line-of-sight velocity dispersion within 0.1$\re$ as $\vdm$,
even when core radii are small, as observations suggest they are for
elliptical galaxies.
However our results also indicated the following
inequalities:

$\ba{cccccl}
1.16 & \leq & \frac{\vdm}{\vlos} & \leq & 1.27 & {\hskip 0.2in} {\rm at}\; R = 0.4\re \\
1.20 & \leq & \frac{\vdm}{\vlos} & \leq & 1.30 & {\hskip 0.2in} {\rm at}\; R = 0.54\re\\
1.24 & \leq & \frac{\vdm}{\vlos} & \leq & 1.37 & {\hskip 0.2in} {\rm at}\; R = \re
\ea$\\

These indicate that the intrinsic scatter of $\vdm$ will be less 
than 10 per cent if we can
measure the line-of-sight velocity dispersion at $\re$ or half $\re$. Then, we
can multiply this velocity dispersion by the average value (e.g. 1.25 for
$R$ = $0.54\re$), in order to get $\vdm$. This argument is
almost  independent of the core radius 
of each galaxy and the anisotropy parameter $\beta$. 

More important, we demonstrated 
that measuring the central velocity dispersion of 
a galaxy can give a misleading representation of $\vdm$, and in particular
can give an overestimate of $\vdm$ which can lead to a higher probability 
for lensing, and hence an inappropriately stringent bound on the cosmological
constant.

Finally, the above estimates are for the case of a purely finite isothermal
distribution.  If one adds to this distribution some central mass, such as
a large central black hole, this will further increase the central velocity
dispersion, and also change the relationship between $\vdm$ and the
line-of-sight velocity dispersion at $\re$ and half $\re$.  By measuring the
velocity dispersion at both these points, however, one can hope to extract out
the central mass contribution and also the isothermal 
contribution \cite{cheng}.
(Alternately, it is clear that the velocity dispersion at $\re$ will be less
sensitive to the former contribution, and thus can be used to approximate the
isothermal contribution.) The important point here is that the mean impact
parameter for lensing at redshifts greater than unity is of order 1-10 kpc.  At
this range a central  mass, if it is of the order of
$\approx 10^9 M_{\sun}$, contributes for example
at the 5-10 percent level to the squared 
velocity dispersion, but it contributes
merely at the 2-5 percent level to the
bend angle for lensing
by the galaxy at this distance 
(the bend angle at some radius $r$ is $(2/\pi)$ times smaller for a point mass
which has the same mass 
as an isothermal sphere at this radius, and hence produces the
same velocity dispersion at this radius, because in the latter case 
the mass outside this radius contributes
to the bend angle but not the velocity dispersion.)
Since the
fourth power 
of the velocity dispersion enters into the optical depth, interpreting
a 10 per cent increase in the squared velocity dispersion using an isothermal
sphere model
predicts a 20 per cent increase
in the optical depth, whereas the actual increase, if the contribution
is from a central compact mass, is less than 10 per cent. 
Hence 
one must be concerned, 
when using velocity dispersions to estimate 
optical depths, about to what extent a possible central mass
contributes to the former but not the latter.

To summarize these arguments one more time: {\it The measured central velocity
dispersion is not necessarily directly correlated to the quantity which is
relevant for lensing statistics.} This fact must be taken into account when
attempting to utilize such statistics to constrain cosmological parameters. 

The last step in the determination of $\sigma_{\rm DM}^*$ is the determination
of the value of $L^*$ and the relation between $L$ and $\vdm$.  We consider
the former first.  
First, to be consistent, it is important to calculate the luminosity function
in the same band as one estimates velocity dispersions, and in which one
performs lensing searches.  Second, it is important to consider not the
luminosity function for all galaxies, but rather for E/S0 galaxies, as
these dominate in the lensing statistics.   An analysis of the latter has been
carried out by Loveday et al. \cite{Loveday}, and yields, in the $\bj$ 
band the relation, for the Schechter function:

\begin{equation}
\alpha = 0.20  ; {\rm M}^* = -19.71 + 5 \log_{10}h
\end{equation}
In order to attempt to confirm this relationship in the $\bt(0)$ band we 
calculated \cite{cheng} a luminosity function utilizing data from a subset
of sample in Faber et al.~\cite{7s} and 
de Vaucouleurs et al.~\cite{RC3}.  In this case, we find a similar relation

\begin{equation}
\alpha = 0.15 \pm 0.55 ; \mbstar = -19.66 + 5 \log_{10}h \pm 0.30
\end{equation}

Note that this best fit M$^*$ differs from that used by Kochanek
\cite{Kochanek96}.
At the same time we stress a fact emphasized by Kochanek, there is
a complicated  interplay in the determination of both
$M^*$ and $\vdm^*$, so that  one cannot arbitrarily vary one without the
other.  In any case, we utilize our relation above for the $\bt(0)$ luminosity
function in what follows below, although we also consider how our results would 
change if the parameters of the luminosity function change.

In order to estimate the velocity dispersion of $L^*$ galaxies we found 42
galaxies with available velocity dispersions, 38 of which were
appropriate for use in our analysis at $R = \re$, and 39 of which had reliable
velocity measures at $R=0.54\re$
\cite{cheng}.  When several different authors differed in their estimates of
either $\re$ or
$\sigma_{\re}$ we utilized the weighted mean of the different estimates and
incorporated the  scatter in our error estimate.  
We then calculated the least-square-fit of the
relation between $\lll$ and $\lv_{\re}$/km s$^{-1})$ \cite{cheng}
and obtained
\be
\lll = (-4.04 \pm 0.49) + (1.89 \pm 0.22)\lv_{\re}/{\rm km\ s}^{-1})
\label{eq:vre}
\ee

If we set $L = L^{*}$ in equation~(\ref{eq:vre}), then we find the velocity
dispersion of luminous elliptical galaxies at the effective radius,
$\vsre = 135.9 \pm 15$ km s$^{-1}$. From the discussion above, we then
multiply this number by 1.31, to get $\vdm^{*} \approx 1.31\vsre \approx$
178 km s$^{-1}$. 

Several issues are relevant to this result.  In the first place, note that the
$L$ vs $\sigma$ relation in the above equation differs from the standard
Faber-Jackson relation~\cite{FJ}.
Note however that this FJ relation is appropriate for
central velocity dispersions.  In addition, it may be there case that the
elliptical galaxies do not form a uniform population, but rather that bright
galaxies and faint galaxies follow separate Faber-Jackson curves (this point
was raised to us by J. Peebles, who has been investigating this issue
\cite{peeblesfuk}). It thus may not be appropriate to enforce this relation on
the bulk sample.  To explore this latter possibility, and because it is
galaxies with the largest velocity dispersions which will dominate in the
analysis of lensing statistics, we considered a subset of our galaxy sample
with $\lv_{\re}$/km s$^{-1}) > 2.2$ and rederived the $L$-$\sigma$ relation.   
In this case, we found:

\be
\lll = (-6.20 \pm 1.82) + (2.83 \pm 0.79)\lv_{\re}/{\rm km\ s}^{-1})
\label{eq:vre2.2}
\ee 
In this case, we find a somewhat higher value of $\vdm^*
=203$ km s$^{-1}$ as might be expected given the features of this subsample. 
It is worth noting that we have used the technique described earlier of using
line-of-sight velocity dispersions at two values of $R$ to extract out the
asymptotic velocity dispersion which may be most appropriate for lensing, and
find, coincidentally that for this subsample a value $\vdm^* = 179$ km
s$^{-1}$.  Nevertheless, as a probe of the robustness of the gravitational
lensing constraints we incorporate both of the above relations as
well as the direct inferred values of $\vdm^*$ in our analysis. Note also that
increasing the slope in the $L$ vs $\vdm$ relation has the same effect on
lensing as  decreasing $\alpha$ in the Schechter function, thus once again
demonstrating the interdependence of these quantities in the deriving
constraints.

Finally, having derived two estimates for $\vdm^*$ here, we must note that the
lower value $\vdm^{*} \approx  178$ km s$^{-1}$ is smaller than that 
utilized in several previous lensing statistical analyses.  This may be due in
part to what we claim is the inappropriate use of central velocity dispersions
in other analyses, and while we reiterate that the value of
$\vdm^*$ cannot be independently varied in a self consistent analysis, and
depends upon the form of the luminosity function, in order to allow for large
possible systematic errors, we also used  $\vdm^{*} \approx 
207$ km s$^{-1}$, which corresponds to a $2\sigma$ variation in the fit to
$\vdm^*$ from first
relation (42) above. Last, for  purposes of comparison with
Kochanek~\cite{Kochanek96}, we considered an  even larger $\vdm^*$ value,
while at the same time adopting his choice of values for the Schechter function
$\alpha =-1$, $L$ vs $\sigma$ slope of 4, (and setting $\rc = 0$ in this case,
see below) in order to explore the effects of variation in choice of $M^*$ and
sample selection.

\section{Review of core radius}

The final quantity which we must determine in order to carry out our analysis is
the relationship between the core radius of
E/S0  galaxy and its luminosity. In Table~\ref{table:lauer}, we list the
inferred  values of core radii from the Class I data in Lauer
\cite{Lauer}, distances of galaxies from Faber et al. \cite{7s}, and
apparent magnitudes ($\bt(0)$) from de Vaucouleurs et al. \cite{RC3}.
With $\mbstar$ = -19.66 + $5 \cdot\log_{10} h \pm 0.30$, 
$\log_{10}(\llstar)$ for each galaxy in Table~\ref{table:lauer}
can be calculated. We plot
$\log_{10}(\llstar)$ against
$\log_{10}(\rc/(h^{-1}$pc)) in  Figure~\ref{fig:lauer}. 
With 13 samples in Table~\ref{table:lauer}, we have found the 
best fit is $\frac{\rc}{\rc^{*}}= (\llstar)^{1.65}$ with 
$\rc^{*}=45\,h^{-1}$pc and $\chi^{2}=11$. If we consider $\Delta \chi^2$ at the 
95 per cent confidence level with three parameters 
of interest (i.e., $\rc^{*}, L^{*}$, and a power law relation between $\rc$ and
$L$), such that the total $\chi^2 = 18.8$ (which is the 95 per cent confidence 
contour line), with the power law relation and $L^{*}$ fixed, then we have 
$\rc^{*}=45\,{\displaystyle_{-17}^{+25}} h^{-1}$pc.
The best fit value of $\rc^{*}$ would in fact agree with an earlier
investigation, if we neglect uncertainties from distances of galaxy samples,
and normalize 
$\mbstar$ to the value chosen in Krauss \& White \cite{KW}.

An interesting test of this fit is to estimate the
$\rc$ of NGC~7457, which is an S0 galaxy in Lauer et al. \cite{Lauer91}. 
Its apparent magnitude ($\bt(0)$) is 11.76 from de Vaucouleurs et al. 
\cite{RC3}, and its distance is about
10 $h^{-1}$Mpc. Based on the formula in this section, we obtain
$\rc = 5.2\,h^{-1}$pc for the best fit of $\rc^{*}$, and the 95 per cent 
confidence lower limit is $\rc = 3.2\,h^{-1}$pc. This result is
marginally consistent  with the lack of evidence for a core radius of NGC~7457
at a limit of $\approx 3.4$ pc ($\hnot = 80$ km s$^{-1}$ Mpc$^{-1}$)
\cite{Lauer91}.

In our analysis,
with the parameters given above and later, without including magnification
bias, we have found that the lensing 
probability is reduced due to 
the core radius by a factor 2 to 3, depending on the value of $\om$ in flat 
cosmological models.  With the magnification bias included, this reduction
factor will be significantly decreased \cite{Kochanek96}.  However, one 
important
result of our analysis is that these effects need not cancel out completely, 
depending upon parameters of the luminosity function, so
that as we shall see, the introduction of a finite core can in fact suppress the
optical depth for lensing, even when magnification bias is taken into account.

\section{Quasar samples and lensing events}

\subsection{Quasar samples}

We first utilized the Large Bright Quasar Survey (LBQS) catalogue 
of Hewett, Foltz \& Chaffe \cite{LBQS}. There are 1055 quasar 
samples in that
catalogue. All quasars are between redshift 0.2 and 3.4, and their apparent
magnitudes ($\bj$) are between 16 and 18.9.  We include these 
flux limits to calculate the
magnification bias in our statistics. There was only one lensing event,
LBQS1009-0252, observed in this survey.  

As a comparison, we also used the
Snapshot survey by Maoz et al.
\cite{snapshot} combined with other surveys by 
Crampton, McClure \& Fletcher \cite{Crampton}, 
Surdej et al. \cite{Surdej}, and Yee, Filippenko \& Tang \cite{Yee}.
There
are 648 quasars in this total sample set. There were 4 lensing events in these
surveys.
In modelling the magnification bias for this set we utilized the flux limit of
19.5, appropriate for the Snapshot survey, which contains the largest number of
quasars, slightly more than the Surdej et al. survey. Our results do not depend
sensitively on this choice, however, and in any case the flux limit in the
latter survey is similar to that in the Snapshot survey.

Finally, if we 
consider all five surveys together, then we will have 1615 quasars in
total, with five lensing
events.  In our analysis of this combined sample we again utilized the 
flux limit from the Snapshot survey (lower than that for the LBQS survey) to
calculate the magnification bias.

\subsection{Lensing events}
Kochanek~\cite{Kochanek93,Kochanek96} has previously discussed the 
rationale for considering the 5 strong
gravitational lensing events mentioned above among all optical gravitational
lensing candidates for use in the calculation of lensing statistics.
We list the quasar redshifts, lens 
redshifts, and angular splittings for these events in Table~\ref{table:events} 
\cite{Q0142,H1413,Q1208,Q1009,rs,PG1115,lens1208}. In 
order to apply our statistical model, we 
actually need to know the redshift of each lens. At present, only 
the redshifts of lenses in two systems (Q0142-100 and PG1115+080) have been 
firmly measured. For those lenses which lack galaxy redshift information,
we chose the most likely absorption lines as lens redshifts. For 
systems H1413+117 and Q1208+1011, we used two possible redshifts in our 
analysis, as two different lenses are possible candidates.

With the information in Table~\ref{table:events}, we can calculate 
the likelihood function in equation~(\ref{eq:Nevent}). What we need are the core
radius and velocity dispersion for each galaxy. Recall the angular separation
is well approximated by $\dtheta=\abtw$ \cite{darkL}. Using the results of
section 3, we employ 
power law relations between luminosity and velocity dispersion as either
$\llstar = (\sigstardm)^{\gamma}$, with (1) $\gamma = 1.89, \vdm^{*} = 178$ km, 
s$^{-1}$ or (2) $\gamma = 2.83, \vdm^{*} = 203$ km s$^{-1}$. 
This implies a power law relation between core radius and
velocity dispersion, i.e., $\frac{\rc}{\rc^{*}} = (\sigstardm)^{3.12}$ or
$\frac{\rc}{\rc^{*}} = (\sigstardm)^{4.67}$.
Combining this equation with a known $\dtheta$ value, and with a given
cosmological model, in principle, we can solve $\rc$ and $\vdm$ for
each lensing galaxy. Actually, except for the lensing galaxy for the system
Q1208+1011, all other four lensing systems allow a second
solution with
$\vdm$ at least 1400 km s$^{-1}$, which is not a suitable scale for a
galaxy. In  our analysis we select the smaller but
reasonable velocity  dispersion. In
Tables~\ref{table:fit_flat}  and~\ref{table:fit_open}, we list our theoretical
predictions of $\vdm$,
$\rc$, and $\bt(0)$ for the lensing galaxy for each lensing system in different
cosmological models.  Note that the inferred $\vdm$ values are reasonable,
but have an average which exceeds the value of $\vdm^*$ we have used.  This is
to be expected.  Lensing galaxies will have preferentially larger velocity
dispersions than the mean, simply because these galaxies are weighted more
heavily in the probability function.  It would thus be inappropriate to use the
mean for lensing galaxies to model all galaxies.

For the five events considered here, we find that 
that 
$\sigstardm$ (or $\llstar$) increases when $\om$ increases for all five events
in flat universe models. This can be explained as follows:
$y_{os}/(y_{ol} y_{ls})$
increases when $\om$ increases for a fixed source redshift and a fixed lens
redshift~\cite{KW,ffkt}. For a fixed (observed) $\dtheta$, when
$y_{os}/(y_{ol} y_{ls})$ increases, $\sigstardm$ has to increase in order to 
prevent $\beta$ from approaching 1/2.
The $\sigstardm$ ratio as a function of $\om$ seems to 
increase faster when $\dtheta$ is smaller. In fact, there is no solution for
$\rc$ and $\vdm$ of the system Q1208+1011 when $\om \geq 0.95$, if $z_l=2.9157$
is chosen. This is because $\yy$ decreases when $\om$ increases for a fixed
source redshift and a fixed lens redshift. As a result, $\dtheta$ from the
system Q1208+1011 is too `large' in the $\om=1$ universe model if the given
source redshift and lens redshift are accurate and correct. If more `large'
angular splitting lensing systems are observed at high redshifts, this will
more strongly favor a low matter density universe, although of course the
number of lensing events will play an important role in constraining models.  In
open universe models, the results are similar to flat  universe models, but
the $\sigstardm$ ratio as a function of $\om$ is smoother.

\section{Numerical results}

\subsection{Magnification bias}
With the parameters discussed above, we can calculate the lensing
optical depth. One interesting task is to compare our bias factor to the SIS
model in Fukugita \& Turner \cite{FT}. We define the average of $B(<m)$ as 
following:
\be
\Bave \;\equiv \frac{1}{\tau} \int B(<m) d\tau
\ee
where $\tau$ is the optical depth of the lensing events. In addition to the
parameters discussed previously, We choose $\mmin$ to be 16, $\mo = 19.15$, 
$a=0.86, b=0.28$ from Hartwick \& Schade \cite{HS}. In 
Figure~\ref{fig:bias}, we have plotted
$\Bave$ as a function of quasar (source) redshift in different
cosmological models, with $m$ = 18.9, 19.5, and 22. 

Although we set a lower limit for the integrals in equation~(\ref{eq:avebias}),
there is no significant impact if we choose $\mmin$ to be either 
16 or smaller (such as negative infinity as in Fukugita \& Turner
\cite{FT}). In Figure~\ref{fig:bias}, we have shown that $\Bave$ is almost
a constant  at higher source redshift ($z_s \geq 1$) in each cosmological
model.  Although the
optical depth of the $\omda = 1$ flat universe model is larger than the 
optical depth of the $\om = 1$ universe model, the averaged total 
magnification bias goes in the other direction. This is also true in the open 
universe model. Comparing to the bias values in Fukugita \& Turner 
\cite{FT},
we have $\Bave$ around 13 when $m=18.9$, but they 
have 7.33. We have $\Bave$ around 7, but they have $B(<m)=4.25$,
when $m = 19.5$. We have $\Bave$ around 4, and they have 
$B(<m)=$ 2.63, when $m=22$. These numbers confirm that although adding a core
radius in the  gravitational lensing analysis will generally reduce the lensing
probability by a factor 2 to 3, the magnification bias including the
core radius is about 1.6 times (depending on the flux limit of the quasar
survey) larger than the SIS model. This is due to the fact that with a non zero
core radius the minimum value of the total amplification of all quasar images is
generally larger than 2, the value for the SIS model.

\subsection{Flat universe models}

Incorporating the magnification bias results discussed above we first consider
the statistical analysis in which quasars from all 5 surveys were utilized. 
For flat universe models, these are presented in columns 2 to 7 in
Table~\ref{table:flat}. As mentioned earlier, the apparent magnitude of
the combined quasar survey limit has been set to be 19.5. If we consider
our best fit galaxy parameters it is clear from
columns 2 to 5 that the expected number of lenses is a good fit, and the
likelihood function is minimized, when
$\om
\approx 0.2$. The strict 95 per cent confidence
level (for 1 degree of freedom) on this quantity for these galaxy parameters
ranges from 
$0.07 < \om < 0.55$ (column 3) or $0.07 < \om < 0.65$ (column5).
However, if we consider the intrinsic scatter merely in velocity dispersions and
core radii, then we find a different result. For example, consider a
smaller $\rc^{*}$, 28 $h^{-1}$pc, and a larger $\sigma^{*}$, 207~km s$^{-1}$. 
Columns 6 and 7 show a best fit around $\om=0.5$, and the 95 per
cent confidence level is $\om > 0.24$. 
This result here is comparable, although less stringent than the 
result (with
$\om=1$ as the  best fit value) given in Kochanek~\cite{Kochanek96},
without including core radii effects.  This demonstrates the sensitivity of 
existing statistical
constraints to the values of the assumed velocity dispersions and  core
radii.  Note that the inclusion of core radii, and the difference in the
assumed value of $M^*$ presumably account for the reduction in the minimum
allowed value of $\om$ compared to Kochanek~\cite{Kochanek96}.  Also note 
that inclusion of core radii is
largely responsible for shifting the peak of the likelihood function in this
case toward values lower than $\om=1$. 
In any case, while suggesting that $\om < 1$ is
preferred by our best fit parameters, the divergence in results demonstrates
the sensitivity of existing statistical constraints to the values of the assumed
velocity dispersions and  core radii. The likelihood functions mentioned above
are plotted in Figure~\ref{fig:likeli19.5}.

Setting $\om=0.2$ in Figure~\ref{fig:number_flat0.2_tot}, we plot the expected 
number of lensing events and the observed (five) events 
vs source redshift ($z_{s}$) . The shape of the predicted 
curve is in agreement with the observation, given the very limited statistics.
If one had a complete quasar survey up to a high redshift (larger than
2.5), one could use this shape function to probe more clearly the agreement, or
lack thereof, between theory and observation. 

In columns 8 to 13 in Table~\ref{table:flat}, we list results obtained by using
only the LBQS survey and the single lens candidate, LBQS1009-0252. Because only
one lens is considered, the results more strongly favor a higher 
$\om$ universe. Nevertheless, while the best fit of $\om$ is around 0.8, the
95 per cent confidence level lower limit is about $\om > 0.15$. The large width
of the allowed region presumably reflects both the reduced statistics, and the
effect of non-zero core radii.  The likelihood functions are plotted in
Figure~\ref{fig:likeli18.9}.  Columns 12 and 13 involve the parameters  
used in Kochanek
\cite{Kochanek96} (i.e.,
$\alpha = -1$,
$\gamma = 4$, 
$\rc^* = 0$, $\sigma^* = 225$ km s$^{-1}$).  The best 
fit of $\om$ from the likelihood analysis is 0.95, with 95 per cent confidence 
level, $\om > 0.2$ in flat universe models (see also 
Figure~\ref{fig:likeli18.9}).  This result is consistent with the result given
by Kochanek~\cite{Kochanek96}, but again the lower limit on $\om$ is
reduced. 

For comparison, in columns 14 to 16 the result using the four combined quasar
surveys \cite{snapshot,Crampton,Surdej,Yee} without LBQS are shown.  As
expected this tends to favor slightly smaller $\om$ than the five survey
analysis. The best fit of $\om$ is about 0.15. 
The 95 per cent confidence levels are $0.03 < \om < 0.43$ and 
$0.03 < \om < 0.38$ for the last two columns.
Statistical values are not available at high $\om$ values in the last two 
columns, because $\vdm$ and $\rc$ of the lensing galaxy in 
system Q1208+1011 cannot be fitted, as discussed in the previous section.

As another probe of the galaxy lensing parameters, if we define the mean angular
splitting as
\be
<\dtheta> \equiv \frac{1}{\tau} \int \dtheta d\tau
\label{eq:dtheta}
\ee
with $\dtheta = \abtw$ \cite{darkL}, we find that $<\dtheta> \approx
2\farcs 9$ ($\gamma=1.89, \vdm^*=178$ km s$^{-1}$) or $2\farcs 0$
($\gamma=2.83, \vdm^*=203$ km s$^{-1}$), when the source redshift is less than 
5 in flat universe models.
However, if we calculate the mean value of the angular separation from
Table~\ref{table:events}, we find a mean value $1\farcs 542$. From the 
formula of $<\dtheta>$ for the SIS model \cite{Kochanek93}, one
can  reduce the mean 
angular splitting by either decreasing $\sigma^*$ or increasing the value of
the parameter $\gamma$ (for fixed Schechter $\alpha$). 
Indeed, if we choose $\vdm^*=178$ km s$^{-1}$ and $\gamma=2.83$, the
combination suggested by extracting away the central mass contribution from the
asymptotic velocity dispersion as discussed previously, we find
$<\dtheta> \approx 1\farcs 5$.  From
Figure~\ref{fig:likeli19.5}, one can infer in this case a best fit for
$\om$  around 0.23, with a 95 per cent
confidence interval around $0.07 < \om < 0.6$.

\subsection{Open universe models}

We repeat the above analysis for open universe models.  When all 5 surveys are
included in the analysis the results are displayed in columns 2 to 5 in
Table~\ref{table:open}. In this case the best fit is $\om \approx 0$.  Note,
that the expected number of lensing events is smaller than the observed
number when one considers more realistic values
$\om > 0$ in this case. The 95 per cent confidence limits are nevertheless
fairly broad, with either
$\om < 0.65$ (column 3) or $\om \leq 1$ (column 5). 

We next consider just the four quasar surveys (Crampton et al.
\cite{Crampton}, Maoz et al. \cite{snapshot}, Surdej et al.
\cite{Surdej}, and Yee et al. \cite{Yee}). The results
are listed in columns 6, 7, and 8 in Table~\ref{table:open}. The best fit value
is again
$\om=0$, with a 95 percent confidence limit of $\om < 0.3$ in this case.

Finally, if we consider the LBQS survey alone, with a single lensed event, it is
clear that higher density open universe models will fare better. Since in fact
the best fit value of $\om$ is in this case close to 0.95, the open universe
model likelihood function in this region is then very similar to that for the
flat model.  As a result, the likelihood function and predicted number of events
will have the same behavior (near $\om =1$) as the values shown in columns 8-13
of Table~\ref{table:flat} and displayed in Figure~\ref{fig:likeli18.9} for the
flat universe case.  As a result, we do not explicitly re-display the values
again in Table~\ref{table:open}.

In general open universe models tend
to predict a redshift distribution for the lenses which is in poorer agreement
with observation, although again because of the very limited statistics this is
not a quantitative problem at this point.  It is also worth noting that one can
increase the number of lensing events by increasing $\vdm^*$ as seen in  column
4, but in this case the mean predicted angular splitting will be increased, as
we have  discussed in the previous subsection. 

\section{Conclusions}

The various different fits described here in 
Tables~\ref{table:flat} and~\ref{table:open} demonstrate that the constraints
one derives on cosmological models from existing lensing statistics depend
in detail on the parameters one uses to model galaxy distributions.  
Nevertheless, the model parameters which we suggest are favored involve a
low density, cosmological constant dominated universe---the model which
coincidentally is favored by other astrophysical data at the present time.  In
our analysis, we have been careful to self consistently obtain the various
lensing parameters (velocity dispersions, core radius, magnification bias,
etc.). As a result, the general features of our analysis are expected to
be robust, even though the intrinsic scatter in values of the selection
functions, galaxy velocity dispersions and galaxy core radii may be large at
the present time. Although the selection function in this paper is as
simple as in Fukugita \& Turner \cite{FT}, a real selection function should
at least account for the image separation and flux ratio. If we
include these factors in our paper, they will reduce the number of predicted
lensing events. With the observed number of lensing events remaining
unchanged, this means that the lensing statistics will favor
a flat universe with a larger cosmological constant value---in even
greater disagreement with Kochanek's conclusion \cite{Kochanek96}.
Quantitatively our analysis suggests a best fit value of
$\om$ ranges between 0.25-0.55 in a flat universe if five lensing
events in the 5 optical surveys considered here are incorporated.  (Note
however that this value can raise to 0.95 if only one lensing event and a
single survey are included in the analysis.)  The distribution of predicted
events is in a good agreement with the distribution of observed events in
source redshift space (see Figure~\ref{fig:number_flat0.2_tot}). Note however
that the results depend on which lensing surveys are utilized and thus the
number of lenses used.  Clearly what is required in order to distinguish
between model parameters are greater statistics from a complete full sky 
quasar survey up to high redshift.  

The predicted number of lenses in open universe models do not vary as sharply
as flat universe models do over the range of $\om$, so while the preferred
value of
$\om$ is around 0 if lensing events from all five lensing surveys considered
here are incorporated, the allowed range remains very broad. The fact that open
universe models tend to predict too few events when compared to 5 lensing
events in the five surveys has also been demonstrated by  Chiba
\& Yoshii
\cite{cy}. Note that even if a larger $\vdm^*$ is chosen compared to those
utilized here, in order to increase the number of lenses, then the mean angular
splitting will become larger than the mean value of the observations in this
case.  Finally, open universe models tend to predict more lower redshift lenses
than flat universe models, and this may be a useful discriminant in the future.

We remind readers that the lensing optical depth is proportional
to the fourth power of the velocity dispersion. A 10 per cent change of velocity
dispersion can cause 46 per cent change in the optical depth. Thus, better
estimates of velocity dispersions for E/S0 galaxies are necessary before the
statistical limits can improve.  We have argued here that determining the
appropriate velocity dispersion for use in lensing statistics, when core radii
are not zero can be somewhat subtle.  Self consistent modelling of the
gravitational potentials may be necessary before line-of-sight velocity
dispersions can be used to infer dark matter velocity dispersions.  In
addition, the effect of compact central mass conglomerations may bias the
interpretation on extracts from using central velocity dispersions in the
derivation of lensing statistics.

The parameter $\gamma$, relating velocity dispersion and luminosity is also
important in gravitational lensing studies. A larger value of $\gamma$
 reduces the mean angular splitting and also pushes $\om$ toward a
low value and a narrow 95 per cent confidence level. (These effects can be
easily understood from equations~(\ref{eq:explens}) and~(\ref{eq:Nevent}).)
Therefore, improved observations of E/S0 galaxy velocity dispersions along
with luminosity measurements in order 
to reduce the intrinsic scatter in the data as well as to better model the 
Schechter parameters would be
useful. 

We have also demonstrated that adding a core radius in the lensing analysis
will generally reduce the total optical depth, even though the magnification
bias factor will be higher than that predicted SIS models.
The overall reduction of the lensing optical depth for non zero core radii is
parameter dependent, however. For example, if we choose a steeper power law
relation between the velocity dispersion and the luminosity ($L \propto
\sigma^4$), with  Schechter
$\alpha= -1$, then including a core radius in the analysis produces no
significant  change in the number of expected lensing events, and produces
shifts in the best fit likelihood analysis for $\om$ of $\le 0.1$. With a
shallower power law relation and the Schechter parameters  used in this paper,
we have however found a 30 per cent to 40 per cent change of the number of
predicted lensing events by adding a core radius. This result is easily
understood: a shallow power law effectively more strongly weights larger
core radii when
$L$ is less than
$L^*$.  

Although the apparent magnitudes are given in $\bt(0)$ band without 
K-corrections in Tables~\ref{table:fit_flat} and ~\ref{table:fit_open}, 
these values nevertheless suggest why observers have not yet found the lensing
galaxy in systems H1413+117, LBQS1009-0252, and Q1208+1011. We also suggest
examining system Q1208+1011 carefully, because this event, 
because of its angular splitting and assumed the high lens redshift,
helps drive the fits to a low $\om$ universe.  Note that recent studies of the
faint lens image of H1413+117  suggest a value of the lens brightness
consistent with that calculated in  Table~\ref{table:fit_flat}
\cite{Turnshek,Kneib}.

To summarize: the self consistent incorporation of core radii in galaxy models,
and fitting the distribution of galaxies to the E/S0 galaxies which
dominate lensing statistics combine together to suggest a statistical best fit
of predictions to the results of existing optical quasar lensing surveys for
a flat universe model with
$\om$ in the range 0.25-0.55.  Considerable systematic uncertainty persists
however in the appropriate velocity dispersion to use in models, as well as the
luminosity-velocity dispersion relation, and the determination of the
appropriate luminosity function parameters.   In addition, how to appropriately
combine the results of different surveys by different groups is not obvious. 
Since the results depend upon all these factors, it may be premature to argue
definitively in favor of this best-fit scenario, although it is encouraging
that it agrees with that obtained using other cosmological observables.  In any
case, we have derived here several new results here associated with
magnification biasing and galaxy models with core radii, as well as the
determination of the appropriate velocity dispersion to use in galaxy lensing
models.  As data improves, we expect that the techniques described here will be
useful in further constraining theoretical models.

\section*{Acknowledgments}

This research work has been partially supported by the Industrial Physics
Group in the Physics Department at Case Western Reserve University, and by
a grant from DOE.
This research has also made use of the NASA/IPAC Extragalactic Database (NED)
which is operated by the Jet Propulsion Laboratory, California Institute
of Technology, under contract with the National Aeronautics and Space
Administration.  We thank Jim Peebles for discussions, and informing us
of his work in progress.

\newpage

\bc\
\epsfxsize=4.0in
\epsffile{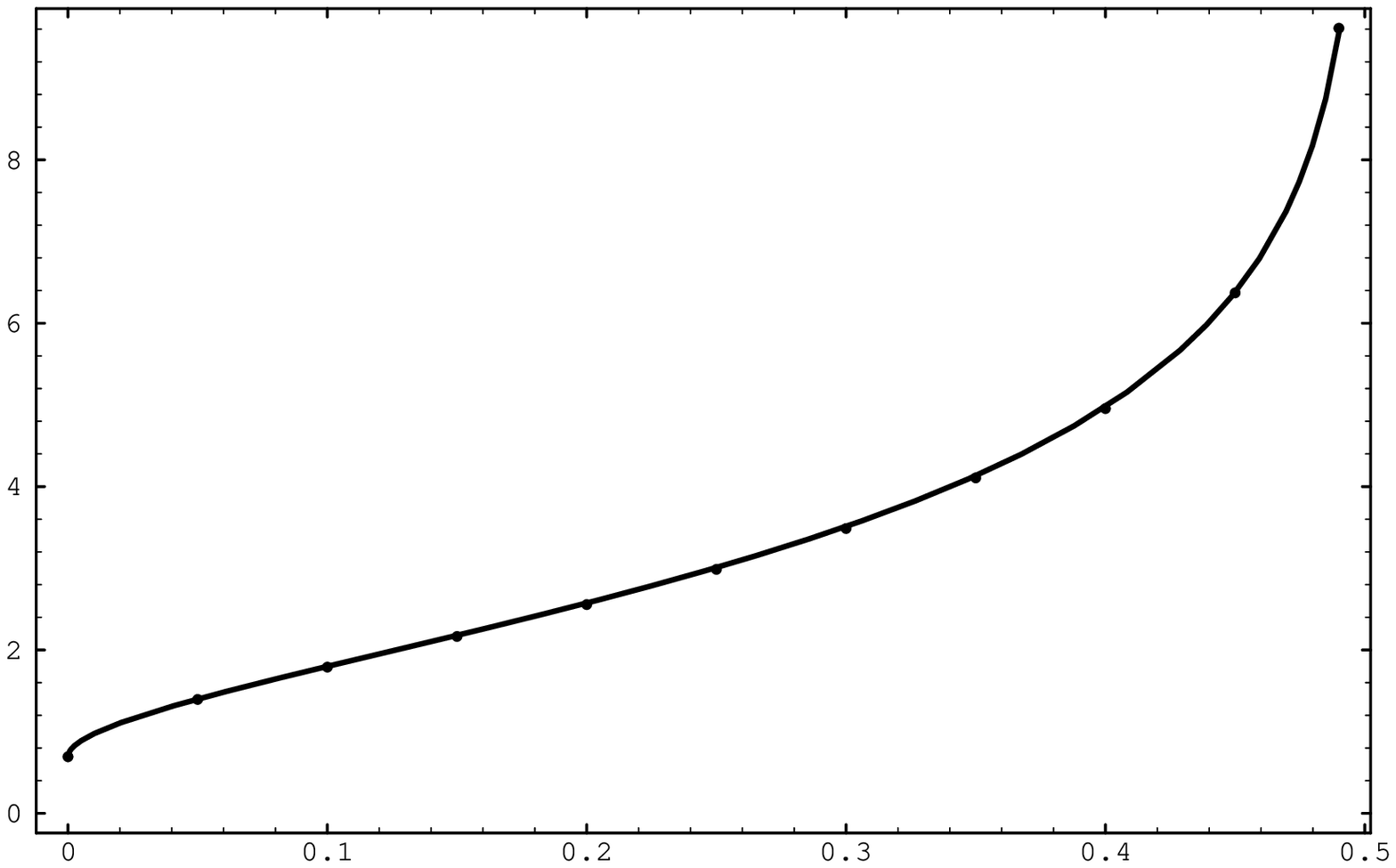}
\ec
\vskip -1.8in
{\hskip 1.0in \rotate{ln($\am$)}}
\vskip 1.1in
\bc
{$\beta$}
\bfe
\caption
{Plot of natural logarithm of $\am$ versus parameter $\beta$. The dots 
are numerical results, and the solid curve is the best fit curve: 
2/$(\lco^{0.65})$.}
\label{fig:amin}
\efe
\ec
 
\bc\
\epsfxsize=4in
\epsffile{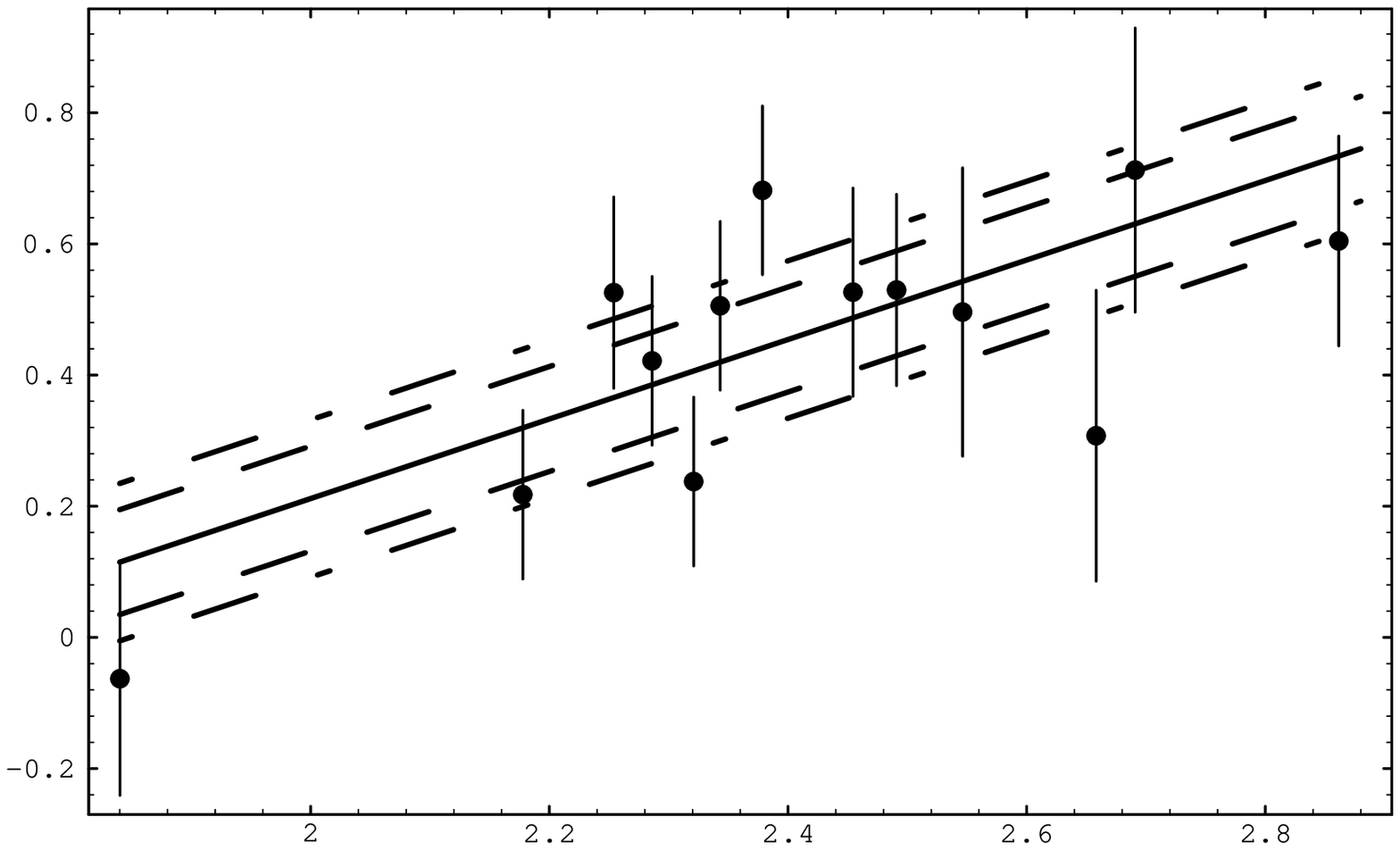}
\ec
\vskip -1.6in
{\hskip 1.0in \rotate[l]{$\log_{10}\llstar$}}
\vskip 0.2in
{\hskip 3.0in $\chi^{2}$ = 11}
\vskip 0.5in
\bc
{$\log_{10}(\rc/(h^{-1}$pc))}
\ec
\bfe
\caption
{The best fit plot of $\log_{10}\llstar$ vs $\log_{10}
(\rc/(h^{-1}$pc)). There are 13 samples with error bars shown in the figure.
The middle solid line is the best fit, and the two dash lines give the 68 per 
cent confidence level, assuming
the best fit power law. The two dash-dot lines 
give the 95 per cent confidence level under the same condition.}
\label{fig:lauer}
\efe

\bc\
\epsfxsize=4in
\epsffile{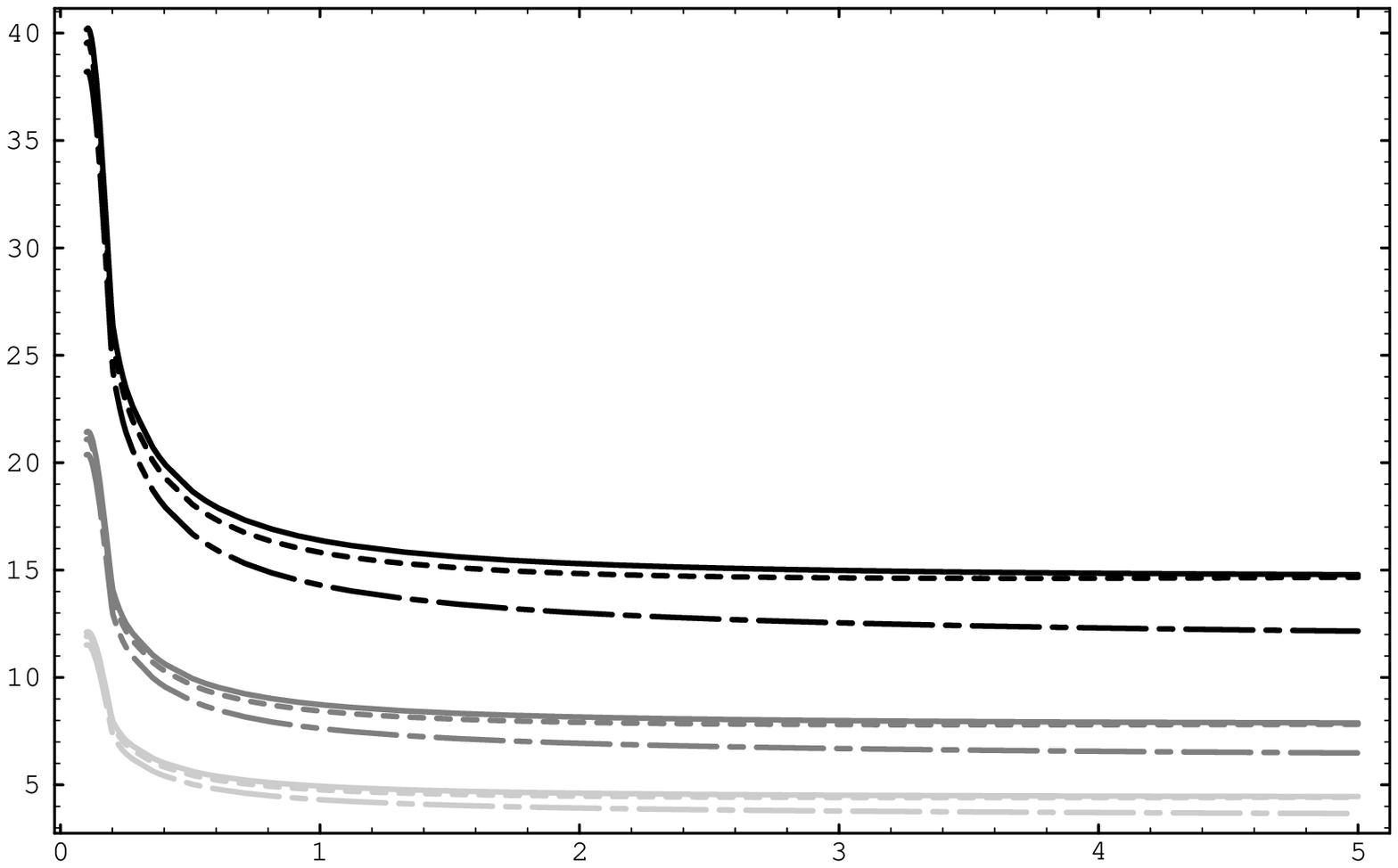}
\ec
\vskip -1.8in
{\hskip 1.0in \rotate[l]{$\Bave$}}
\vskip 0.7in
\bc
{$z_s$}
\bfe
\caption
{Plots of $\Bave$ versus quasar (source) redshift. The solid curve
sets are for a flat universe model, with $\om=1$. The dash-dot curve sets
are also for a flat universe model, but with $\om=0$, and $\omda=1$. The
dot-dot curve sets are for an open universe model, with $\om=0$ and zero
cosmological constant. The black (highest) curves represent quasar survey
limit, $m = 18.9$. The gray curves represent quasar survey limit, $m = 
19.5$. The light gray (lowest) curves represent quasar survey limit, $m = 22$.
We choose $\vdm^*=178$ km s$^{-1}$, $\rc^*=45 h^{-1}$ pc, and $\gamma=1.89$ to 
generate these curves.}
\label{fig:bias}
\efe
\ec

\bc\
\epsfxsize=4in
\epsffile{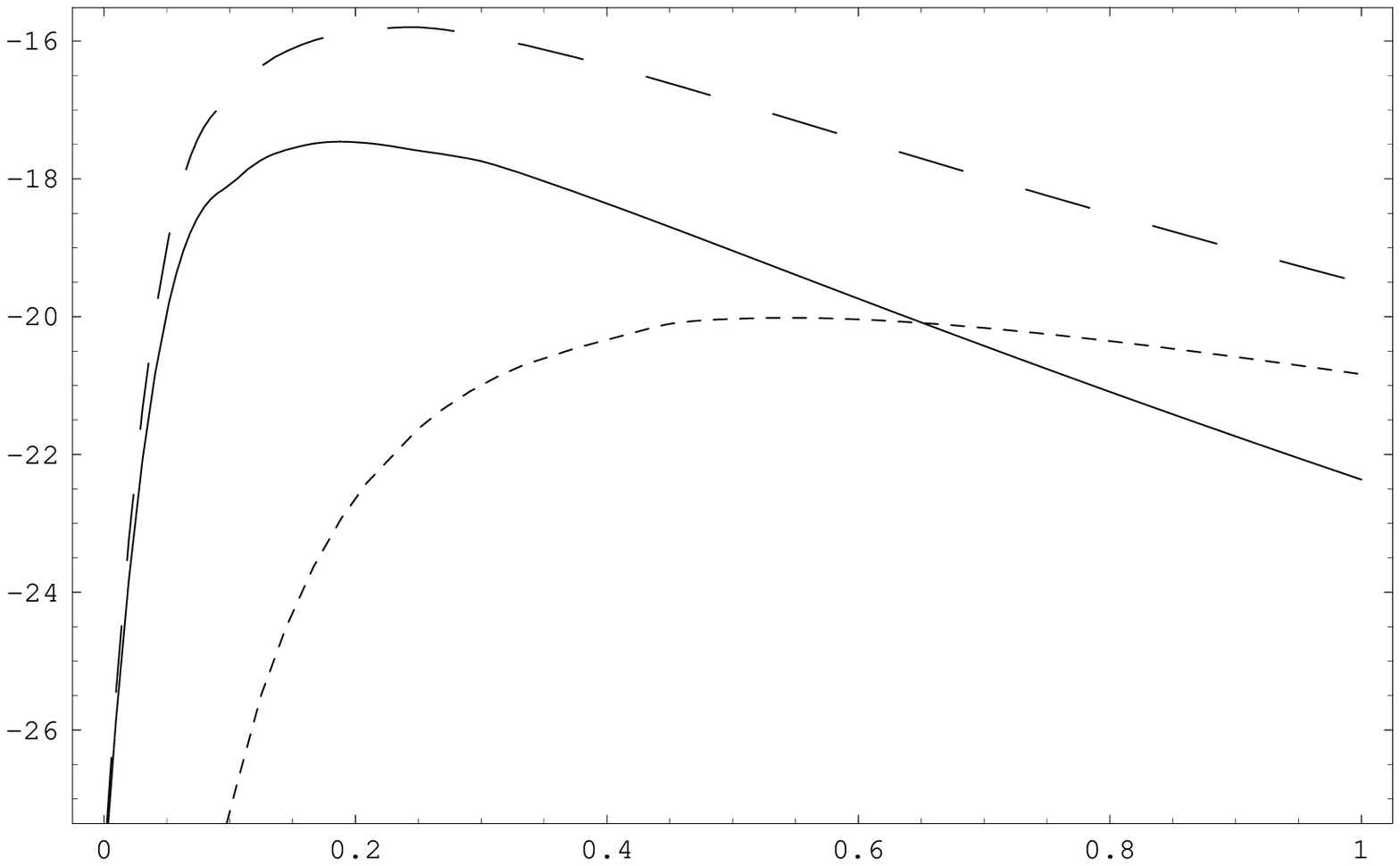}
\ec
\vskip -1.7in
{\hskip 1.0in \rotate[l]{$\ln \cal L$}}
\vskip 1.2in
\bc
{$\om$}
\bfe
\caption
{Plots of likelihood functions vs $\om$. The solid curve is the plot of
the third column in Table~\protect\ref{table:flat}, 
with $\rc^{*}=$ 45 $h^{-1}$pc, $\vdm^{*}=$ 203~km s$^{-1}$, and
$\gamma=2.83$. The dash-dash curve represents the result for 
$\rc^{*}=$ 45 $h^{-1}$pc, $\vdm^{*}=$ 178~km s$^{-1}$, and $\gamma=1.89$. The
dot-dot curve is the plot for $\rc^{*}=$ 28 $h^{-1}$pc, $\vdm^{*}=$ 207~km
s$^{-1}$, and $\gamma=1.89$.}
\label{fig:likeli19.5}
\efe
\ec

\bc\
\epsfxsize=4in
\epsffile{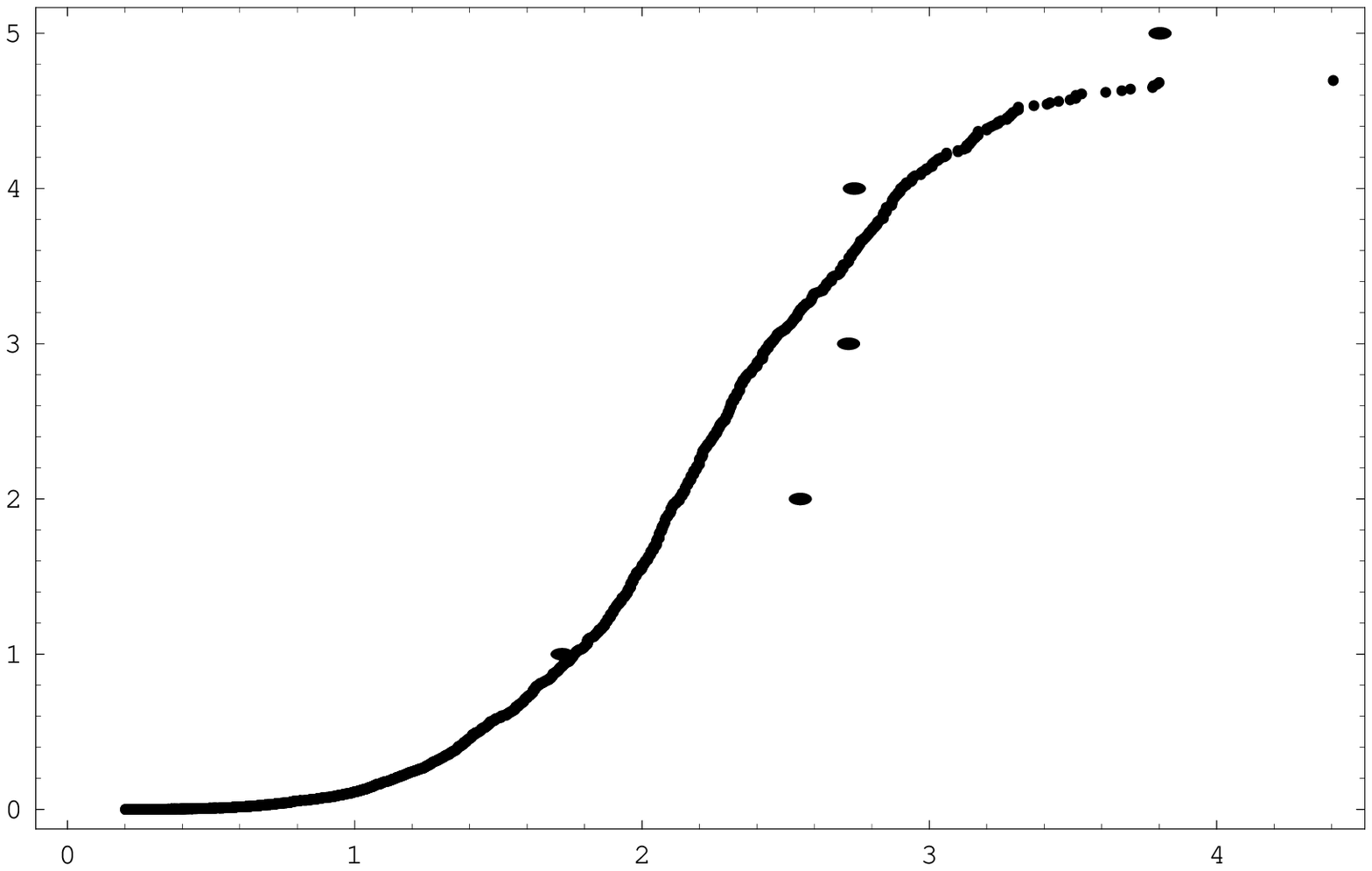}
\ec
\vskip -1.9in
{\hskip 1.0in \rotate[l]{$N(<z_{s})$}}
\vskip 1.0in
\bc
{$z_s$}
\bfe
\caption
{The predicted number of events and observed events are plotted
as a function of quasar redshifts from all 5 quasar surveys with $\om=0.2$ in a flat 
universe model. The small dots making up the curve give the total 
predicted number of lensing events within a given source redshift. The 
ellipses give the number of 
observed events. The parameters are $\vdm^*=203$ km s$^{-1}$,
$\rc^*=45 h^{-1}$ pc, and $\gamma=2.83$.}
\label{fig:number_flat0.2_tot}
\efe
\ec

\bc\
\epsfxsize=4in
\epsffile{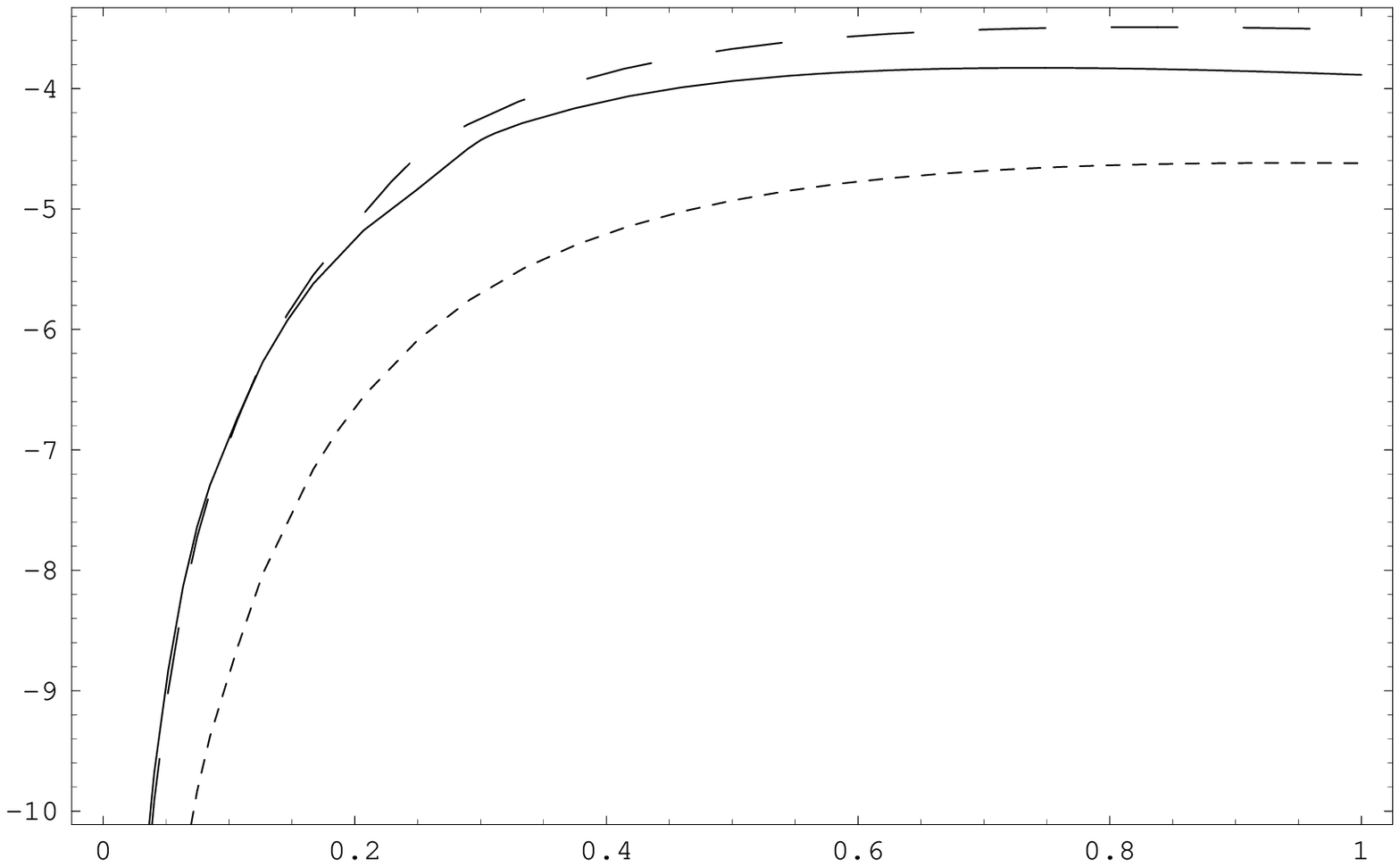}
\ec
\vskip -1.7in
{\hskip 1.0in \rotate[l]{$\ln \cal L$}}
\vskip 1.1in
\bc
{$\om$}
\bfe
\caption
{Plots of likelihood functions vs $\om$. The solid curve is the plot for
$\rc^{*}=$ 45 $h^{-1}$pc, $\vdm^{*} =$ 203~km s$^{-1}$, and $\gamma=2.83$
in flat universe models using only the LBQS result. The
dash-dash curve represents the plot for $\rc^{*}=$ 45 $h^{-1}$pc, $\vdm^{*}=$
178~km s$^{-1}$, and $\gamma=1.89$. The dot-dot curve is the plot for
$\rc^{*}=0$, $\vdm^{*}=$ 203~km s$^{-1}$, $\gamma=4$ and luminosity function
parameter $\alpha =-1$.}
\label{fig:likeli18.9}
\efe
\ec

\bc
\btr{ccccccccccc} \hline
$\beta$ & 0 & 0.025 & 0.05 & 0.075 & 0.1 & 0.125 & 0.15 & 0.175 & 0.2 & 0.225\\
$<\!A\!>$ & 4 & 5.81 & 7.08 & 8.44 & 9.99 & 11.9 & 14.1 & 16.9 & 20.4 & 24.9\\
\hline
\etr
\vskip 12pt
\btr{cccccccccc} \hline
$\beta$ & 0.25 & 0.275 & 0.3 & 0.325 & 0.35 & 0.375 & 0.4 & 0.425 & 0.5\\ 
$<\!A\!>$ & 30.7 & 38.7 & 49.9 & 66.3 & 91.6 & 134 & 213 & 383 & $\infty$ \\
\hline
\etr
\begin{table}[h]
\caption{Averaged amplification as a function of $\beta$.}
\label{table:ampave}
\end{table}
\ec

\bc
\btr{llll} \hline
$\beta$ & $\am$    & $L_1(\am)$  &   $\lco$    \\ \hline
0       & 2        & 1           &   1         \\
0.05    & 4.03321  & 0.538516    &   0.580948  \\
0.1     & 5.98959  & 0.381634    &   0.427036  \\
0.15    & 8.71889  & 0.276166    &   0.318752  \\
0.2     & 12.8853  & 0.19801     &   0.235198  \\
0.25    & 19.8151  & 0.137874    &   0.168375  \\
0.3     & 32.6748  & 0.0910677   &   0.114322  \\
0.35    & 60.7457  & 0.0549151   &   0.0708916 \\
0.4     & 141.863  & 0.0278588   &   0.037016  \\
0.45    & 585.195  & 0.00920176  &   0.0126015 \\
0.49    & 14931.9  & 0.000780135 &   0.001096  \\ 
0.5	& $\infty$ & 0		 &   0	       \\ \hline
\etr
\begin{table}[h]
\caption{$\am$, $L_1(\am)$, and $\lco$ as a function of $\beta$.}
\label{table:amin}
\end{table}
\ec

\bc
\btr{cccc} \hline
NGC  & $\theta_{c}$ (arcsec) & Distance ($h^{-1}$Mpc) & $m(\bt(0))$ \\ \hline
720  & 4.58$\pm$0.01 & 20.5$\pm$4.4 & 11.13 \\
741  & 1.91$\pm$0.03 & 53$\pm$11 & 12.18 \\
1395 & 1.86$\pm$0.12 & 19.9$\pm$1.9 & 10.52 \\
1407 & 3.21$\pm$0.07 & 19.9$\pm$1.9 & 10.51 \\
1600 & 3.73$\pm$0.09 & 40.2$\pm$4.9 & 11.85 \\
3379 & 1.66$\pm$0.07 & 8.6$\pm$1.3 & 10.17 \\
4261 & 2.61$\pm$0.05 & 27.8$\pm$5.9 & 11.32 \\
4365 & 2.33$\pm$0.02 & 13.33$\pm$0.71 & 10.42 \\
4374 & 2.99$\pm$0.02 & 13.33$\pm$0.71 & 9.91 \\
4472 & 3.7$\pm$0.03 & 13.33$\pm$0.71 & 9.26 \\
4636 & 3.24$\pm$0.17 & 13.33$\pm$0.71 & 10.37 \\
4649 & 3.41$\pm$0.11 & 13.33$\pm$0.71 & 9.7 \\
5846 & 2.51$\pm$0.14 & 23.4$\pm$2.8 & 10.87 \\ \hline
\etr
\begin{table}[h]
\caption{Core radii of sample galaxies and their corresponding distances and
apparent magnitudes.}
\label{table:lauer}
\end{table}
\ec

\bc
\btr{lcccc} \hline
Source        & $z_{s}$ & $z_{l}$ & $m_{l}$  & $\dtheta$ (arcsec) \\ \hline
Q0142-100     & 2.719   & 0.49    & R $\approx$ 19   & 2\farcs 2 \\
PG1115+080    & 1.722   & 0.294   & R $\approx$ 19.8 & 2\farcs 3 \\
H1413+117     & 2.551   & 1.4382$^{a}$ (1.6603$^{a}$) & & 1\farcs 23 \\
LBQS1009-0252 & 2.739   & 0.869$^{a}$ & R $>$ 21 & 1\farcs 53 \\ 
Q1208+1011    & 3.803   & 1.1349$^{a}$ (2.9157$^{a}$) &      & 0\farcs 45 \\ \hline
\multicolumn{5}{p{4.9in}}{$^a$ The assumed galaxy redshifts, based on the 
most possible absorption lines by observers in literature.}\\ \hline
\etr
\begin{table}[h]
\caption
{Parameters of gravitational lensing systems. First column: the name of each 
quasar in each lensing system. Second column: the source redshift. Third 
column: the lens redshift. Fourth column:
the apparent magnitude (in R band) of each galaxy, if the value is available.
Fifth column: the angular splitting of each lensing system. We take the
average separation value of the quadruple system H1413+117.}
\label{table:events}
\end{table}
\ec

\bc
\btr{lcccccc} \hline
 & \multicolumn{3}{c}{$\om$=0, $\omda$=1} & \multicolumn{3}{c}
{$\om$=0.2, $\omda$=0.8}\\ \hline
source        & $\vdm$(km s$^{-1}$) & $\rc(h^{-1}$pc) & $\bt(0)$ 
& $\vdm$(km s$^{-1}$) & $\rc(h^{-1}$pc) & $\bt(0)$  \\ \hline
Q0142-100     & 217 & 62  & 21.8 & 233 & 86 & 21.4 \\ 
PG1115+080    & 221 & 67  & 20.4 & 230 & 81 & 20.1 \\ 
H1413+117     & 223 & 69  & 25.2 & 275 & 185 & 23.9 \\ 
H1413+117$^a$ & 250 & 119 & 25.3 & 324 & 398 & 23.8 \\ 
LBQS1009-0252 & 198 & 40  & 23.9 & 224 & 72  & 23.1 \\ 
Q1208+1011    & 106 & 2.1 & 26.7 & 124 & 4.5 & 25.7 \\ \hline
\multicolumn{7}{p{5.5in}}{$^a$ Galaxy redshift is 1.66.} \\ \hline
\etr
\begin{table}[h]
\caption
{Theoretical predictions of velocity dispersions, core radii,
and apparent magnitudes of lenses in flat universe models. K-correction is not 
included in the fourth and seventh columns. We choose $\vdm^*=203$ km s$^{-1}$, 
$\rc^* =45 h^{-1}$ pc, and $\gamma=2.83$.}
\label{table:fit_flat}
\end{table}
\ec

\newpage

\bc
\btr{lcccccc} \hline
 & \multicolumn{3}{c}{$\om$=0, $\omr$=1} & \multicolumn{3}{c}
{$\om$=0.25, $\omr$=0.75}\\ \hline
source        & $\vdm$(km s$^{-1}$) & $\rc(h^{-1}$pc) & $\bt(0)$ 
& $\vdm$(km s$^{-1}$) & $\rc(h^{-1}$pc) & $\bt(0)$  \\ \hline
Q0142-100     & 254 & 129 & 21.0 & 254 & 127 & 20.9 \\ 
PG1115+080    & 244 & 107 & 19.8 & 244 & 107 & 19.8 \\
H1413+117     & 316 & 354 & 23.3 & 320 & 376 & 23.1 \\
H1413+117$^a$ & 390 & 954 & 23.1 & 420 & 1338 & 22.7 \\
LBQS1009-0252 & 252 & 123 & 22.5 & 251 & 122 & 22.4 \\
Q1208+1011    & 141 & 8.3 & 25.1 & 140 & 8.0 & 24.9 \\ \hline
\multicolumn{7}{p{5.5in}}{$^a$ Galaxy redshift is 1.66.}\\ \hline
\etr
\begin{table}[h]
\caption
{Theoretical predictions of velocity dispersions, core radii,
and apparent magnitudes of lenses in open universe models. K-correction is not 
included in the fourth and seventh columns. We choose $\vdm^*=203$ km s$^{-1}$, 
$\rc^* =45 h^{-1}$ pc, and $\gamma=2.83$.}
\label{table:fit_open}
\end{table}
\ec

\newpage

\begin{table}
\rotate[l]{\begin{minipage}{9in}
\btr{lccccccccccccccc}\hline
$\om$ &  $\nexp^{a,b}$ & $\ln {\cal L}^{a,b}$ & $\nexp^{a,c}$ & 
$\ln {\cal L}^{a,c}$ & $\nexp^{a,d}$ & $\ln {\cal L}^{a,d}$ & 
$\nexp^{b,e}$ & $\ln {\cal L}^{b,e}$ & $\nexp^{c,e}$ & $\ln {\cal L}^{c,e}$ &
$\nexp^{e,f}$ & $\ln {\cal L}^{e,f}$ & $\nexp^{c,g}$ & $\ln {\cal L}^{c,g}$ &
$\ln {\cal L}^{c,g,h}$ \\ \hline
0 & 21.0 & -28.16 &	22.2 & -27.95 &	47.7 & -54.69 &	14.7 & -14.91 &	15.5 & -15.49 &	19.5 & -19.16 &	16.1 & -20.63 & -20.73 \\
0.05 & 10.8 & -19.93 &	11.3 & -18.98 &	24.8 & -33.43 &	8.3 & -8.95 &	8.8 & -9.13 &	11.3 & -11.53 &	7.7 & -15.37 & -15.75 \\
0.1 & 7.5 & -18.11 &	7.9 & -16.86 &	17.4 & -27.25 &	6.0 & -6.92 &	6.3 & -6.95 &	8.3 & -8.88 &	5.2 & -14.57 & -15.12 \\
0.15 & 5.8 & -17.56 &	6.1 & -16.11 &	13.5 & -24.30 &	4.7 & -5.87 &	5.0 & -5.81 &	6.6 & -7.50 &	4.0 & -14.55 & -15.20 \\
0.2 & 4.7 & -17.47 &	4.9 & -15.87 &	11.0 & -22.63 &	3.9 & -5.24 &	4.1 & -5.13 &	5.4 & -6.65 &	3.2 & -14.78 & -15.52 \\
0.25 & 3.9 & -17.59 &	4.1 & -15.80 &	9.3 & -21.63 &	3.3 & -4.83 &	3.5 & -4.56 &	4.7 & -6.09 &	2.7 & -15.04 & -15.85 \\
0.3 & 3.4 & -17.74 &	3.6 & -15.93 &	8.0 & -20.99 &	2.9 & -4.43 &	3.0 & -4.25 &	4.1 & -5.70 &	2.3 & -15.43 & -16.29 \\
0.35 & 3.0 & -18.03 &	3.1 & -16.12 &	7.1 & -20.59 &	2.5 & -4.24 &	2.6 & -4.03 &	3.6 & -5.41 &	2.0 & -15.84 & -16.75 \\
0.4 & 2.6 & -18.36 &	2.8 & -16.36 &	6.3 & -20.34 &	2.3 & -4.10 &	2.4 & -3.87 &	3.2 & -5.20 &	1.7 & -16.25 & -17.22 \\
0.45 & 2.4 & -18.70 &	2.5 & -16.61 &	5.7 & -20.10 &	2.0 & -4.01 &	2.1 & -3.76 &	2.9 & -5.05 &	1.6 & -16.67 & -17.68 \\
0.5 & 2.1 & -19.04 &	2.2 & -16.88 &	5.1 & -20.03 &	1.8 & -3.94 &	1.9 & -3.67 &	2.7 & -4.93 &	1.4 & -17.08 & -18.14 \\
0.55 & 2.0 & -19.39 &	2.0 & -17.16 &	4.7 & -20.02 &	1.7 & -3.89 &	1.8 & -3.61 &	2.5 & -4.84 &	1.3 & -17.50 & -18.60 \\
0.6 & 1.8 & -19.74 &	1.9 & -17.43 &	4.3 & -20.04 &	1.6 & -3.86 &	1.6 & -3.56 &	2.3 & -4.77 &	1.2 & -17.92 & -19.06 \\
0.65 & 1.7 & -20.08 &	1.7 & -17.70 &	4.0 & -20.09 &	1.4 & -3.84 &	1.5 & -3.53 &	2.1 & -4.72 &	1.1 & -18.35 & -19.52 \\
0.7 & 1.5 & -20.42 &	1.6 & -17.98 &	3.7 & -20.16 &	1.3 & -3.83 &	1.4 & -3.51 &	2.0 & -4.68 &	1.0 & -18.78 & -19.99 \\
0.75 & 1.4 & -20.76 &	1.5 & -18.24 &	3.5 & -20.25 &	1.3 & -3.83 &	1.3 & -3.50 &	1.8 & -4.66 &	0.9 & -19.22 & -20.48 \\
0.8 & 1.3 & -21.09 &	1.4 & -18.51 &	3.2 & -20.35 &	1.2 & -3.83 &	1.2 & -3.49 &	1.7 & -4.64 &	0.9 & -19.70 & -20.99 \\
0.85 & 1.2 & -21.42 &	1.3 & -18.76 &	3.0 & -20.46 &	1.1 & -3.84 &	1.1 & -3.49 &	1.6 & -4.63 &	0.8 & -20.22 & -21.54 \\
0.9 & 1.2 & -21.74 &	1.2 & -19.02 &	2.9 & -20.58 &	1.0 & -3.85 &	1.1 & -3.49 &	1.5 & -4.62 &	0.7 & -20.85 & -22.21 \\
0.95 & 1.1 & -22.05 &	1.1 & -19.26 &	2.7 & -20.71 &	1.0 & -3.87 &	1.0 & -3.50 &	1.5 & -4.62 &	0.7 & N/A & N/A \\
1. & 1.0 & -22.36 &	1.1 & -19.51 &	2.6 & -20.84 &	0.9 & -3.89 &	1.0 & -3.51 &	1.4 & -4.62 &	0.7 & N/A & N/A \\ \hline
\multicolumn{16}{p{6.5in}}{$^a$ All five quasar surveys and lensing events are used.}\\
\multicolumn{16}{p{6.5in}}{$^b$ $\vdm^* =203$ km s$^{-1}$, $\rc^*=45 h^{-1}$ pc, and $\gamma=2.83$.}\\
\multicolumn{16}{p{6.5in}}{$^c$ $\vdm^* =178$ km s$^{-1}$, $\rc^*=45 h^{-1}$ pc, and $\gamma=1.89$.}\\
\multicolumn{16}{p{6.5in}}{$^d$ $\vdm^* =207$ km s$^{-1}$, $\rc^*=28 h^{-1}$ pc, and $\gamma=1.89$.}\\
\multicolumn{16}{p{6.5in}}{$^e$ Only LBQS and LBQS1009-0252 are considered.}\\
\multicolumn{16}{p{6.5in}}{$^f$ $\vdm^* =225$ km s$^{-1}$, $\rc^*=0$, $\gamma=4$, and $\alpha=-1$.}\\
\multicolumn{16}{p{6.5in}}{$^g$ Based on surveys by Crampton et al.
\cite{Crampton}, Maoz et al. \cite{snapshot}, Surdej et al.
\cite{Surdej}, and Yee et al. \cite{Yee}. Four lensing events
are used. $z_l=2.9157$ in system Q1208+1011 is used. There is no solution for
$\rc$ and $\vdm$ of the system Q1208+1011 when $\om \geq 0.95.$} \\
\multicolumn{16}{p{6.5in}}{$^h$ $z_{l}=1.66$ in system H1413+117.}\\ \hline
\etr
\caption
{Expected lensing events and likelihood analysis. This table is for flat
universe models, i.e., $\om+\omda=1$.
$\nexp$ is our theoretical prediction of the number of lensing events, and
$\ln {\cal L}$ is the logarithm of the maximum likelihood
function. $z_{l}=1.438$ in system H1413+117 and $z_l=1.1349$ in system
Q1208+1011 are generally used, unless otherwise noted.}
\label{table:flat}
\end{minipage}}
\end{table}

\clearpage
\newpage

\bc
\btr{lccccccc@{\hspace{2in}}l} \hline
$\om$ &  $\nexp^{a,b}$ & $\ln {\cal L}^{a,b}$ & $\nexp^{a,c}$ &
$\ln {\cal L}^{a,c}$ & $\nexp^{d}$ & $\ln {\cal L}^{d}$ &
$\ln {\cal L}^{d,e}$ & \\ \hline
0 & 2.1 & -19.43 &	5.0 & -22.34 &	1.3 & -16.04 & -16.99 & \\
0.05 & 1.9 & -19.61 &	4.7 & -22.38 &	1.3 & -16.40 & -17.39 & \\
0.1 & 1.9 & -19.79 &	4.5 & -22.43 &	1.2 & -16.74 & -17.76 & \\
0.15 & 1.8 & -19.96 &	4.3 & -22.49 &	1.2 & -17.06 & -18.12 & \\
0.2 & 1.7 & -20.06 &	4.1 & -22.49 &	1.1 & -17.29 & -18.38 & \\
0.25 & 1.6 & -20.23 &	4.0 & -22.56 &	1.1 & -17.59 & -18.70 & \\
0.3 & 1.6 & -20.39 &	3.8 & -22.64 &	1.0 & -17.87 & -19.01 & \\
0.35 & 1.5 & -20.55 &	3.7 & -22.72 &	1.0 & -18.14 & -19.31 & \\
0.4 & 1.5 & -20.70 &	3.6 & -22.80 &	0.9 & -18.41 & -19.61 & \\
0.45 & 1.4 & -20.86 &	3.5 & -22.89 &	0.9 & -18.68 & -19.89 & \\
0.5 & 1.4 & -21.01 &	3.3 & -22.98 &	0.9 & -18.94 & -20.18 & \\
0.55 & 1.3 & -21.15 &	3.2 & -23.07 &	0.9 & -19.21 & -20.47 & \\
0.6 & 1.3 & -21.30 &	3.2 & -23.16 &	0.8 & -19.48 & -20.76 & \\
0.65 & 1.3 & -21.44 &	3.1 & -23.25 &	0.8 & -19.75 & -21.06 & \\
0.7 & 1.2 & -21.58 &	3.0 & -23.34 &	0.8 & -20.04 & -21.36 & \\
0.75 & 1.2 & -21.71 &	2.9 & -23.43 &	0.8 & -20.35 & -21.69 & \\
0.8 & 1.2 & -21.85 &	2.8 & -23.52 &	0.7 & -20.69 & -22.06 & \\
0.85 & 1.1 & -21.98 &	2.8 & -23.61 &	0.7 & -21.11 & -22.49 & \\
0.9 & 1.1 & -22.11 &	2.7 & -23.70 &	0.7 & -21.79 & -23.18 & \\
0.95 & 1.1 & -22.24 &	2.6 & -23.79 &	0.7 & N/A & N/A & \\
1. & 1.0 & -22.36 &	2.6 & -23.88 &	0.7 & N/A & N/A & \\ \hline
\multicolumn{8}{p{5.5in}}{$^a$ All five quasar surveys and lensing events are used.}\\
\multicolumn{8}{p{5.5in}}{$^b$ $\vdm^* =203$ km s$^{-1}$, $\rc^*=45 h^{-1}$ pc, and $\gamma=2.83$.}\\
\multicolumn{8}{p{5.5in}}{$^c$ $\vdm^* =207$ km s$^{-1}$, $\rc^*=28 h^{-1}$ pc, and $\gamma=1.89$.}\\
\multicolumn{9}{p{6.4in}}{$^d$ $\vdm^* =178$ km s$^{-1}$, $\rc^*=45 h^{-1}$ pc, and $\gamma=1.89$.
Based on surveys by Crampton et al.
\cite{Crampton}, Maoz et al. \cite{snapshot}, Surdej et al.
\cite{Surdej}, and Yee et al. \cite{Yee}. Four lensing events
are used. $z_l=2.9157$ in system Q1208+1011 is used. There is no solution for
$\rc$ and $\vdm$ of the system Q1208+1011 when $\om \geq 0.95.$}\\
\multicolumn{8}{p{5.5in}}{$^e$ $z_{l}=1.66$ in system H1413+117.}\\ \hline
\etr
\ec
\begin{table}[h]
\caption
{Expected lensing events and likelihood analysis. This table is for open
universe models, i.e., $\omda=0$.
$\nexp$ is our theoretical prediction for the number of lensing events, and
$\ln {\cal L}$ is the logarithm of the maximum likelihood
function. $z_{l}=1.438$ in system H1413+117 and $z_l=1.1349$ in system
Q1208+1011 are generally used, unless otherwise noted.}
\label{table:open}
\end{table}

\end{document}